\documentclass[%
 reprint,
 amsmath,amssymb,
 aps,
pra,
]{revtex4-1}

\usepackage{graphicx}
\usepackage{dcolumn}
\usepackage{bm}
\usepackage[T1]{fontenc}
\usepackage[utf8]{inputenc}
\usepackage{lmodern}
\usepackage{subfigure}
\usepackage{braket}

\newcommand{\bracket}[2]{\langle#1|#2\rangle}

\begin{document}

\preprint{APS/123-QED}

\title{Inversion of coherent backscattering with interacting Bose--Einstein condensates in two--dimensional disorder : a Truncated Wigner approach}

\date{\today}

\author{Renaud Chr\'etien} \email{rchretien@uliege.be}
\author{Peter Schlagheck} 

\affiliation{CESAM Research Unit, University of Liege, 4000 Li\`ege, Belgium}

\begin{abstract}
We theoretically study the propagation of an interacting Bose-Einstein condensate in a two-dimensional disorder potential, following the principle of an atom laser. The constructive interference between time--reversed scattering paths gives rise to coherent backscattering, which may be observed under the form of a sharp cone in the disorder--averaged angular backscattered current. As is found by the numerical integration of the Gross-Pitaevskii equation, this coherent backscattering cone is inversed when a non--vanishing interaction strength is present, indicating a crossover from constructive to destructive interferences. Numerical simulations based on the Truncated Wigner method allow one to go beyond the mean--field approach and show that dephasing renders this signature of antilocalisation hidden behind a structureless and dominant incoherent contribution as the interaction strength is increased and the injected density decreased, in a regime of parameters far away from the mean--field limit. However, despite a partial dephasing, we observe that this weak antilocalisation scenario prevails for finite interaction strengths, opening the way for an experimental observation with $^{87}$Rb atoms.
\end{abstract}

\pacs{Valid PACS appear here}
\maketitle


\section{Introduction}
The phenomenon of coherent backscattering (CBS) \cite{MaretPRL1985,WolfP.E.1988,refId0,PhysRevLett.55.2692} lies at the heart of various interference effects in mesoscopic transport physics. CBS arises due to constructive interferences between a path and its time--reversed counterpart in the backscattered direction that survive ensemble averaging. It is observed in a very wide variety of situations, ranging from the explanation that Saturn's rings are twice brighter \cite{HAPKE2002523} in the backscattered direction to the probing of deep underground to search for oil \cite{Margerin2009}. It has been experimentally verified by illuminating a powder with light \cite{PhysRevLett.55.2692,MaretPRL1985,WolfP.E.1988,WiersmaPRL1995} or for elastic \cite{PhysRevLett.84.1693} and acoustic waves \cite{PhysRevLett.79.3637} and most recently with Bose-Einstein condensates in the presence of a two--dimensional disorder, when time--of--flight imaging has shown that the condensate, initially well--prepared in a momentum state $\mathbf{p}_i$, experiences a momentum redistribution over $2\pi$ with a notable peak in backward direction \cite{PhysRevLett.109.195302}. In the context of electronic transport, coherent backscattering gives rise to weak localisation \cite{PhysRevB.22.5142,BERGMANN19841}. It also inhibits thermalisation and quantum ergodicity \cite{Engl2014PRL} in closed many--body systems and can give rise to many--body spin echoes \cite{EnglPRA2018}. CBS is seen as a precursor of strong (Anderson) localisation \cite{Anderson1958PR,PhysRevB.22.4666} for which coherent forward scattering \cite{CherroretPRL2012,CherroretPRA2014,MiniaturaPRA2014,MuellerPRL2014,WellensPRA2016} has been recently identified as a key indicator.

Weak localisation and coherent backscattering may be affected by nonlinearities \cite{PhysRevLett.100.033902}. Such nonlinearities arise for example in the context of light within nonlinear media \cite{Finlayson1990APL,Hennig1999PR}. In the presence of a nonlinearity, the angular profile of light scattering by a disordered opaque medium displays a narrow dip in the backscattered direction \cite{AgranovichPRB1991}. In the context of ultracold atoms, the mean--field description of the atomic gas accounts for the presence of atom--atom interaction via the nonlinear Gross--Pitaevskii equation. Mean--field studies \cite{Hartung2008PRL,Hartmann2012AP} have indeed shown that in a quasi--steady context, the coherent backscattering peak can be inverted, as a result of the presence of a nonlinearity in the wave equation describing the quantum transport of matter wave towards a disordered region (see also Ref.\ \cite{AgranovichPRB1991}). Similar results have been obtained with Aharonov--Bohm rings in the presence of disorder, where the presence of a nonlinearity gives rise to an inversion of Al'tshuler--Aronov-Spivak oscillations \cite{ChretienPRA18}. 

A natural question that arises in this context is to what extent the mean--field approximation remains valid, particularly concerning the peak inversion. A study based on diagrammatic many--body techniques \cite{Geiger2013NJoP} predicts that a dephasing effect in the presence of strong interaction is expected. Indeed, in the presence of a finite interaction strength, inelastic collision processes that are not described in the mean--field approximation yield an energy redistribution amongst the interacting particles. Owing to that energy redistribution, an incoherent current is produced, which can eclipse the coherent contribution due to interference effects. This was also confirmed by Ref.\ \cite{ScoquartPRR2020} where it was found that in the non-equilibrium configuration of Ref.\ \cite{PhysRevLett.109.195302}, CBS is reduced owing to thermalization--induced dephasing.

In order to obtain a complementary point of view to that issue, we use the truncated Wigner method \cite{Wigner1931,Wigner1932PR,Moyal1949PCPS,Steel1998PRA,Sinatra2002JPBAMOP,Polkovnikov2003PRA} which is a quasiclassical method that allows to go beyond the mean--field regime. The truncated Wigner method takes into account quantum fluctuations by a random sampling of the initial quantum state evolved along Gross--Pitaevskii trajectories. It therefore allows for the description of both coherent and incoherent processes that are not described by the Gross--Pitaevskii equation. Those incoherent processes are of particular relevance in an experimental context, since they can overshadow the interference effect that one wants to highlight in the context of CBS. The truncated Wigner method thus provides a convenient tool to probe in which regime of atom density and interaction dephasing dominates, and can inform about the feasibility of a transport experiment. In this latter context, it was applied to investigate the flow of a Bose--Einstein condensate across obstacles and disorder potentials \cite{ScottPRA2008}, to study atom--laser scenarios \cite{Dujardin2015PRA,DujardinAdP2015}, and it was shown that it provides reliable predictions for average particle densities and currents in disordered systems \cite{Dujardin2016}. In the context of disordered Aharonov--Bohm rings, truncated Wigner simulations predict the inversion of Al'tshuler--Aronov--Spivak oscillations in a finite regime of interaction but also the presence of dephasing for stronger interactions \cite{ChretienPRA18}.

In this paper, we apply the truncated Wigner method in order to study the crossover from constructive to destructive interferences due to the interaction in a 2D Bose--Einstein condensate. For that purpose, we study the same system as in Ref.\ \cite{Hartung2008PRL}, that is, a source of atoms that injects a coherent bosonic matter--wave beam onto a two--dimensional disordered slab of finite width. Our numerical findings confirm the inversion of coherent backscattering within the mean--field regime, and truncated Wigner simulations show that beyond this regime the inversion is partially destroyed, but remains observable in a regime that should be accessible experimentally, before getting hidden by dephasing.

In Sec.\ \ref{sec:sys}, we start by introducing the physical configuration we study. In view of numerically integrating the equations describing the configuration we study, we discuss in Sec.\ \ref{sec:nummeth} the spatial discretisation scheme that we use in this context. We then briefly explain in  the numerical methods we use in this paper, namely the numerical integration of the Gross--Pitaevskii equation, as well as the truncated Wigner method. Sec.\ \ref{sec:res} is devoted to the discussion of the numerical results. We study the occurence of coherent backscattering and its inversion in the mean--field regime, as is documented in \cite{Hartung2008PRL}. We examine the prevalence of this inversion beyond the mean--field regime with the truncated Wigner method and identify in which regime the inverted CBS peak is still visible.

\section{Description of the scattering configuration \label{sec:sys}}
\begin{figure}[h!]
\begin{center}
\includegraphics[width=7cm]{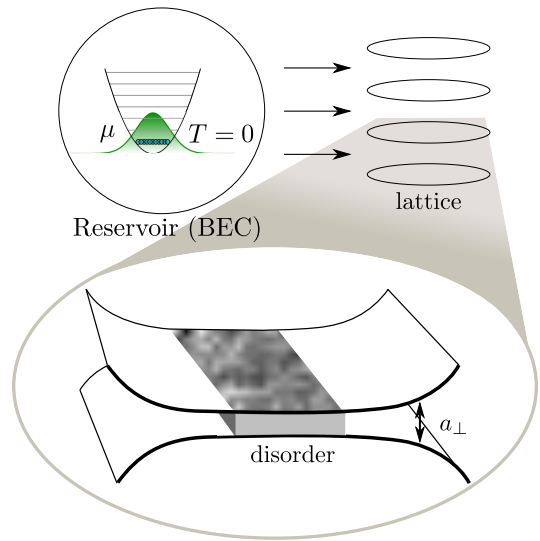}
\end{center}
\caption{Sketch of the scattering configuration. A Bose--Einstein condensate at temperature $T=0$ and chemical potential $\mu$ is injected in a spatially localised disorder potential superimposed by a one--dimensional optical lattice that confines the bosons in quasi two--dimensional planes located between the layers of the lattice.}
\label{fig:sys}
\end{figure}

We consider a Bose--Einstein condensate of $\mathcal{N} \rightarrow \infty$ particles at zero temperature $T = 0$ which is outcoupled from a reservoir with a finite chemical potential $\mu>0$, following the principle of an atom laser \cite{Bloch1999PRL,Guerin2006PRL,Couvert2008EEL,Gattobigio2009PRA,Gattobigio2011PRL,Vermersch2011PRA,Bolpasi2014NJP}. The outcoupled particles are propagating towards a two--dimensional disorder potential $V(\mathbf{r})$ which can be experimentally realised by means of, for instance, optical speckle fields \cite{PRLLye2005}. The confinement to a planar motion in two dimensions can be achieved by superimposing a strong one--dimensional optical lattice in a direction perpendicular to the propagation direction. The incident beam will then be squeezed in a stack of 2D layers and thus propagates quasi two--dimensionally, as is represented in \textsc{Fig}.\ \ref{fig:sys}. 

A many--body model for the description of such a quantum transport problem is given by a set of evolution equations for the field operator $\hat{\psi}(\mathbf{r},t)$ of the bosons in the scattering region, where $\mathbf{r} \equiv (x,y)$ is the spatial position, and for the particle annihilation operator of the source $\hat{\phi}_S(t)$, whose evolution equations are given by \cite{ChretienPRA18,Ernst2010PRA}
\begin{align}
i\hbar \dfrac{\partial \hat{\psi}(\mathbf{r},t)}{\partial t} & = \hat{H}_0 \hat{\psi}(\mathbf{r},t)+ \tilde{g}(\mathbf{r}) \hat{\psi}^\dagger(\mathbf{r},t)\hat{\psi}(\mathbf{r},t)\hat{\psi}(\mathbf{r},t) \nonumber \\
& \hspace{1cm} + K(\mathbf{r},t) \hat{\phi}_S(0)e^{-i\mu t/\hbar}\hat{\phi}_\mathcal{S}(t) \label{eq:eveq1} \\
i\hbar \dfrac{\partial \hat{\phi}_S(t)}{\partial t} & = \mu \hat{\phi}_S(t) + \int \mathrm{d}\mathbf{r} K^*(\mathbf{r},t) \hat{\psi}(\mathbf{r},t). \label{eq:eveq2}
\end{align}
In Eqs. \eqref{eq:eveq1} and \eqref{eq:eveq2}, $\hat{H}_0 = -(\hbar^2/2m)\Delta + V(\mathbf{r}$) is the two--dimensional single--particle Hamiltonian describing the propagation of particles in the disorder potential $V(\mathbf{r})$, with $\hbar$ the reduced Planck constant and $m$ the mass of the atoms. We also introduced the position--dependent coupling strength $K(\mathbf{r},t)$ which couples the reservoir to the scattering region and $\tilde{g}(\mathbf{r})$, the effective interaction strength. In the presence of a 2D confinement, it is given by $\tilde{g}(\mathbf{r}) \equiv \hbar^2 g(x)/m$, with the 2D effective dimensionless interaction strength
\begin{equation}
g(x) = 2\sqrt{2\pi}\frac{a_s}{a_\perp(x)},
\label{eq:g2D}
\end{equation}
where $a_s$ is the s--wave scattering length of the atoms under study and $a_\perp(x) = \sqrt{\hbar/m\omega_\perp(x)}$ the oscillator length associated with the transverse confinement. Due to the spatial dependence of $a_\perp(x)$ which follows the spatial profile of the confining potential, the interaction strength depends on the longitudinal propagation coordinate. We assume it to be adiabatically and smoothly ramped on from zero at position $x_L$ to a finite value $g_\text{max}$ at position $x_L + \Delta x$ and then ramped off from $g_\text{max}$ at position $x_R$ to zero at position $x_R + \Delta x$, using a smooth switching function \cite{HartmannThesis} represented in panel (e) of \textsc{Fig}.\ \ref{fig:dens} whose profile \eqref{eq:func1} is detailed in Appendix \ref{app:timo}.

The reservoir is given by a trapped Bose--Einstein condensate filled with $\mathcal{N} \rightarrow \infty$ atoms. In order to ease the calculations, we choose an idealised profile for the coupling
\begin{equation}
K(\mathbf{r},t) = \kappa(t)\delta(x-x_\mathcal{S})\phi(y),
\label{eq:idealised}
\end{equation}
which acts as a source that injects particles into the scattering plane at position $x_\mathcal{S}$, with a transverse profile $\phi(y) = 1$ that we assume to be homogeneous. The temporal profile of the coupling $\kappa(t)$ is adiabatically and smoothly ramped from zero to a constant value $\mathcal{\kappa}_\text{max}$ (this can be experimentally achieved by varying the intensity of the radio--frequency field in case of outcoupling via a radio--frequency knife \cite{Guerin2006PRL}) following the smooth switching function \cite{HartmannThesis} profile \eqref{eq:func2} described in Appendix \ref{app:timo}. If this coupling tends to zero in such a manner that the product $\mathcal{N}\vert \kappa(t)\vert^2$ remains constant \cite{Guerin2006PRL,Riou2008PRA,Dujardin2014APB}, then a stationary many--body scattering state can be realised. 

The disorder potential we use in this paper is generated by
\begin{equation}
V(\mathbf{r}) = V_0 \int \dfrac{1}{\sqrt{\pi}\sigma} \exp\left(\dfrac{-\vert \mathbf{r} - \mathbf{r}'\vert^2}{2\sigma^2}\right)\eta(\mathbf{r}') \mathrm{d}\mathbf{r}',
\label{eq:dispot}
\end{equation}
where the correlator $\eta(\mathbf{r})$ is chosen as a Gaussian white noise satisfying $\langle \eta(\mathbf{r}) \rangle = 0$ as well as $ \langle \eta(\mathbf{r}) \eta(\mathbf{r}') \rangle = \delta(\mathbf{r} - \mathbf{r}')$. This potential is such that its probability distribution to obtain a certain value for $V$ is given by the gaussian distribution.
\begin{equation}
P(V) = \dfrac{1}{\sqrt{2\pi V_0}} e^{-V^2/(2V_0)^2}.
\end{equation}
The disorder potential in Eq.\ \eqref{eq:dispot} is such that its average value vanishes $\langle V(\mathbf{r})\rangle = 0$ (with $\langle \cdot \rangle$ the random average) and its two--point correlation function 
\begin{equation}
\langle V(\mathbf{r})V(\mathbf{r}')\rangle = V_0^2 \exp\left(\dfrac{-\vert \mathbf{r} - \mathbf{r}'\vert^2}{4\sigma^2}\right)
\end{equation}
is of Gaussian shape.  Note that even though the Gaussian two point correlator is not identical to the one describing an optical speckle field, the predictions obtained with this disorder potential are nevertheless expected to be very similar to the ones that result from a speckle disorder, provided the disorder correlation lengths are identical \cite{Paul2009PRA}.

\section{Numerical methods \label{sec:nummeth}}
\subsection{Discretisation procedure}
In order to numerically implement the truncated Wigner method, we perform a discretisation of the 2D scattering region of length $\mathcal{L}$ and width $\mathcal{W}$ resulting in a series of $L \times W$ sites labelled by $l$ and $w$ and spaced by the grid spacing $\delta$. As is usually done in that case, we describe the kinetic energy operator in terms of a finite--difference scheme
\begin{align}
\dfrac{\partial^2 \hat{\psi}(x,y)}{\partial x^2}  & \simeq \dfrac{\hat{\psi}(x+\delta,y) + \hat{\psi}(x-\delta,y) - 2 \hat{\psi}(x,y)}{\delta^2}, \label{eq:fd1} \\
\dfrac{\partial^2 \hat{\psi}(x,y)}{\partial y^2}  & \simeq \dfrac{\hat{\psi}(x,y+\delta) + \hat{\psi}(x,y-\delta) - 2 \hat{\psi}(x,y)}{\delta^2}. \label{eq:fd2}
\end{align}
As a result of the finite--difference scheme discretisation, each site acquires both an on--site energy $E_{\delta} = \hbar^2/m\delta^2$ and a nearest--neighbour hopping term $E_\delta/2$. The Hamiltonian resulting from the discretisation of space reads
\begin{align}
\hat{H} & = \sum_{l=1}^{L}\sum_{w=1}^{W} \Big[ 2E_{\delta} \hat{a}_{l,w}^\dagger \hat{a}_{l,w} + V_{l,w} \hat{a}_{l,w}^\dagger \hat{a}_{l,w} \nonumber \\
& \hspace{2cm} - \dfrac{E_{\delta}}{2} \left( \hat{a}_{l+1,w}^\dagger \hat{a}_{l,w} + \hat{a}_{l,w}^\dagger\hat{a}_{l+1,w} \right) \nonumber \\
& \hspace{2cm} - \dfrac{E_{\delta}}{2} \left( \hat{a}_{l,w+1}^\dagger \hat{a}_{l,w} + \hat{a}_{l,w}^\dagger\hat{a}_{l,w+1} \right)  \nonumber \\
& \hspace{3cm} + E_\delta g_l \hat{a}_{l,w}^\dagger \hat{a}_{l,w}^\dagger \hat{a}_{l,w} \hat{a}_{l,w}\Big] \nonumber \\
& \hspace{0.4cm} + \sum_{j=1}^{W} \left[\kappa(t)\hat{a}^\dagger_{l_\mathcal{S},j}\hat{b} + \kappa^*(t)\hat{b}^\dagger\hat{a}_{l_\mathcal{S},j}\right] + \mu \hat{b}^\dagger \hat{b},
\label{eq:Ham}
\end{align}
where $\hat{a}_{l,w}^\dagger$ (resp. $\hat{a}_{l,w}$) is the creation (resp. annihilation) operator at site $(l,w)$ and $\hat{b}^\dagger$ (resp. $\hat{b}$) is the creation (resp. annihilation) operator of the source which is maintained at the chemical potential $\mu$ and vanishing temperature $T=0$. We implement smooth exterior complex scaling in the longitudinal direction in order to absorb outgoing waves \cite{Balslev1971CMP,Simon1973AM,Simon1979PLA,Junker1982AAMP,Reinhardt1982ARPC,Ho1983PR,Loewdin1988AQC,Rom1990JCP,Moiseyev1998PR,Moiseyev2005PRA,Dujardin2015PRA,ChretienPRA18} and we consider periodic boundary conditions in the transverse direction. 

The on--site interaction parameter is defined as
\begin{equation}
U(x) = \tilde{g}(x)/\delta^2 = g_lE_\delta = \dfrac{4\pi\hbar^2a_S}{\sqrt{2\pi}m\delta^2 a_\perp(x)}
\label{eq:naichoi}
\end{equation}
and is controlled by the dimensionless parameter $g(x)=2\sqrt{2\pi}a_S/a_\perp(x)$. As Ref.\ \cite{Ron017JPB} indicates, this choice for the on--site interaction parameter exhibits convergence issues in the formal limit $\delta \to 0$. In Appendix \ref{app:renorm}, we determine the correct scaling of this interaction parameter as a function of $\delta$ and conclude that for the choice $k\delta = 1$ we made, corrections to the scaling \eqref{eq:naichoi} are negligible.

In order to properly discretise the disorder potential, we discretise the $\delta$ distribution in the two--point correlation function of the correlator, which amounts to generating within the disordered slab complex Gaussian random numbers $\xi_{l,w}$ which fulfil
\begin{equation}
\langle \xi_{l,w}\xi_{l',w'}\rangle = \delta_{l,l'}\delta_{w,w'}
\end{equation}
and satisfy the periodic boundary conditions $\xi_{l,w+W} = \xi_{l,w}$. One has then to take the convolution product of those numbers with a Gaussian envelope so that the disorder at point $(l,w)$ is generated by
\begin{equation}
V_{l,w} = \sum_{l'=-l_\text{start}}^{l_\text{end}}\sum_{w'=-\infty}^{\infty}A_{l,l'}A_{w,w'}\xi_{l',w'},
\label{eq:disorder}
\end{equation}
with the Gaussian weight 
\begin{equation}
A_{j,j'} = \sqrt{\dfrac{V_0 \delta}{\sqrt{\pi}\sigma}} \exp\left[-\dfrac{\delta^2}{2\sigma^2}\left(j-j'\right)^2\right],
\end{equation}
where $V_0$ is the disorder strength and $\sigma$ its correlation length. 

In the Heisenberg picture, the Hamiltonian provided in Eq.~\eqref{eq:Ham} yields the evolution of the annihilation operators according to
\begin{align}
i\hbar \dfrac{\partial \hat{a}_{l,w}(t)}{\partial t} & = (2E_{\delta} + V_{l,w}) \hat{a}_{l,w}(t) \nonumber \\
& \hspace{0.5cm} - \dfrac{E_{\delta}}{2}\left[\hat{a}_{l-1,w}(t)+\hat{a}_{l+1,w}(t)\right] \nonumber\\
& \hspace{0.5cm} - \dfrac{E_{\delta}}{2}\left[\hat{a}_{l,w-1}(t)+\hat{a}_{l,w+1}(t)\right] \nonumber\\ 
& \hspace{0.5cm} + E_\delta g_{l} \hat{a}_{l,w}^\dagger(t) \hat{a}_{l,w}^2(t) + \kappa(t) \delta_{l,l_{\mathcal{S}}} \hat{b}(t) \label{eq:evop1} \\
i\hbar \dfrac{\partial \hat{b}(t)}{\partial t} & = \mu \hat{b}(t) + \sum_{j=1}^{W}\kappa^*(t)\hat{a}_{l_\mathcal{S},j}(t). \label{eq:evop2}
\end{align}  
In the absence of interaction and disorder, a steady many--body scattering state can be achieved. It is characterised by a stationary density and current that are given by \cite{Dujardin2015PRA,Dujardin2014APB,ChretienPRA18}
\begin{align}
\rho^\varnothing & = \dfrac{1}{\delta^2} \dfrac{\mathcal{N}\vert \kappa(t)\vert^2}{\mu(2E_\delta - \mu)} \\
j^\varnothing & = \dfrac{1}{\hbar} \dfrac{\mathcal{N}\vert \kappa(t)\vert^2}{\sqrt{\mu(2E_\delta - \mu)}}. 
\label{eq:denscurr}
\end{align}

\subsection{Mean--field Gross--Pitaevskii approach}
In the limit of a large atomic density and small interaction strength, the numerical integration of the Gross--Pitaevskii equation has revealed to provide very satisfactory descriptions in various atom--lasers scenarios \cite{Leboeuf2001PRA,Carusotto2001PRA,Paul2005PRL,Paul2005PRA,Paul2007PRA}. The principle of the mean--field approximation lies in the fact that quantum operators can be replaced by c--numbers as long as on-site densities are large and the interaction strength weak. In this approximation, and making the ansätze $\psi_{l,w}(t) = \langle \hat{a}_{l,w} \rangle e^{-i\mu t/\hbar}$ and $\chi(t) = \langle \hat{b} \rangle e^{-i\mu t/\hbar}$, Eqs.\ \eqref{eq:evop1} and \eqref{eq:evop2}
reduce to the discretised Gross--Pitaevskii equation 
\begin{align}
i\hbar \dfrac{\partial \psi_{l,w}(t)}{\partial t} & = (2E_{\delta} + V_{l,w}-\mu) \psi_{l,w}(t) \nonumber \\
& \hspace{0.5cm} - \dfrac{E_{\delta}}{2}\left[\psi_{l-1,w}(t)+\psi_{l+1,w}(t)\right] \nonumber\\
& \hspace{0.5cm} - \dfrac{E_{\delta}}{2}\left[\psi_{l,w-1}(t)+\psi_{l,w+1}(t)\right] \label{eq:GP1} \\ 
& \hspace{0.5cm} + E_\delta g_{l} \vert\psi_{l,w}(t)\vert^2 \psi_{l,w}(t) + \kappa(t) \delta_{l,l_{\mathcal{S}}} \chi(t) \nonumber \\
i\hbar \dfrac{\partial \chi(t)}{\partial t} & = \sum_{w=1}^{W}\kappa^*(t)\psi_{l_\mathcal{S},w}(t),
\label{eq:GP2}
\end{align}
with the initial conditions $\psi_{l,w}(t_0) = 0$ and $\chi(t_0) = \sqrt{\mathcal{N}}$, corresponding to an empty scattering region and a coherent Bose--Einstein condensate within the reservoir of atoms. 

Inspecting Eq.~\eqref{eq:GP1} and \eqref{eq:GP2}, we can deduce that $\chi(t) = \sqrt{\mathcal{N}} \left[1 + \mathcal{O}(\vert\kappa\vert^2)\right]$ for some finite time interval $t-t_0$, implying that in the formal limit where the coupling $\kappa$ tends to zero in such a manner that $\mathcal{N} \vert\kappa\vert^2$ remains constant\footnote{This implies in practice that one would consider a large population of reservoir atoms (say, $\mathcal{N} \sim 10^4$) and a small outcoupling amplitude (say, $\kappa \sim 10^{-2}$ in the natural units that we consider here) in such a way that the two compensate each other, giving rise to a finite product $\mathcal{N} \kappa^2$.}, $\chi(t)$ can be safely assumed to be constant in time, thereby yielding a nonlinear Schrödinger equation with a source term \cite{Paul2005PRL,Paul2007PRA,Ernst2010PRA} given by
\begin{align}
i\hbar \dfrac{\partial \psi_{l,w}(t)}{\partial t} & = \left(\dfrac{2E_{\delta}}{q_{l,w}} - \mu q_{l,w} + V_{l,w}\right) \psi_{l,w}(t) \nonumber \\
& \hspace{0.5cm} - \dfrac{E_{\delta}}{2}\left[J_{l-}\psi_{l-1,w}(t)+J_{l+}\psi_{l+1,w}(t)\right] \nonumber\\
& \hspace{0.5cm} - \dfrac{E_{\delta}}{2}\left[\psi_{l,w-1}(t)+\psi_{l,w+1}(t)\right] \label{eq:SchrSourterm} \\ 
& \hspace{0.5cm} + E_\delta g_{l} \vert\psi_{l,w}(t)\vert^2 \psi_{l,w}(t) + \sqrt{\mathcal{N}} \kappa \delta_{l,l_{\mathcal{S}}} \nonumber
\end{align}
Here, an effective hopping term
\begin{equation}
J_{l\pm} = \left(\dfrac{1}{q_{l\pm 1,w}} - \dfrac{\delta}{2}\dfrac{q_{l \pm 1,w}'}{q_{l \pm 1,w}^2}\right)
\label{eq:cs}
\end{equation}
is introduced to implement complex scaling where, within the scattering region, $q_{l,w} = 1$, leaving the Hamiltonian unchanged, whereas outside the scattering region $q_{l,w}$ is smoothly ramped to $e^{i\theta}$ so that the $x$ coordinate is rotated in the complex plane according to $x \mapsto z = x e^{i\theta}$, with $\theta > 0$ the rotation angle \cite{Balslev1971CMP,Simon1973AM,Simon1979PLA,Junker1982AAMP,Reinhardt1982ARPC,Ho1983PR,Loewdin1988AQC,Rom1990JCP,Moiseyev1998PR,Moiseyev2005PRA,Dujardin2015PRA,ChretienPRA18}. This rotation of the $x$ coordinate allows to absorb outgoing waves and hence to model open systems.

The approach developed here, which has been used in various situations \cite{Leboeuf2001PRA,Carusotto2001PRA,Paul2005PRL,Paul2005PRA,Paul2007PRA}, suffers from a major drawback. Because of two--body scattering \cite{Dujardin2016,Geiger2012PRL,Geiger2013NJoP}, a non--condensed population can be created as a result of a weak atom--atom interaction, particularly in the presence of disordered potentials. Those effects must be tackled by means of a method going beyond the mean--field approach.

\subsection{Truncated Wigner method \label{sec:tw}} 
The drawback related to the effects beyond the mean--field approach can be overcome with the truncated Wigner method \cite{Wigner1931,Wigner1932PR,Moyal1949PCPS,Steel1998PRA,Sinatra2002JPBAMOP,Polkovnikov2003PRA}, which has been successfully used in the context of atom--laser scenarios \cite{Dujardin2015PRA,DujardinAdP2015}. This method consists in finding a map between the von Neumann equation governing the time evolution of the density matrix of the system and the related Wigner function \cite{Wigner1931,Wigner1932PR} $\mathcal{W}(\{\psi_{l,w},\psi_{l,w}^*\},t)$ defined in the phase space spanned with the classical fields $\psi_{l,w}$ at sites $(l,w)$. The resulting equation, containing third order derivatives of the classical fields $\psi_{l,w}$, is practically impossible to integrate because of the prohibitively large dimension of the underlying phase space \cite{Dujardin2015PRA}. The principle of the truncated Wigner method lies in the omission of those third--order derivative terms, hence resulting in a Fokker--Planck equation with a drift term. The former can be mapped to a set of coupled Langevin equations for the time--dependent canonically conjugated variables $\psi_{l,w}(t)$ and $\psi^*_{l,w}(t)$, which we refer to as classical field amplitudes. The evolution equation is given by 
\begin{align}
i\hbar \dfrac{d}{dt} \psi_{l,w} & = \left(\dfrac{2E_{\delta}}{q_{l,w}} - \mu q_{l,w} + V_{l,w}\right) \psi_{l,w} \nonumber \\
& \hspace{1cm} - \dfrac{E_\delta}{2} \left(J_{l+} \psi_{l+1,w} + J_{l-} \psi_{l-1,w}\right) \nonumber \\
& \hspace{1cm} - \dfrac{E_\delta}{2}\left(\psi_{l,w+1} + \psi_{l,w-1}\right) \nonumber \\
& \hspace{1cm} + E_\delta g_l (\vert\psi_{l,w}\vert^2-1)\psi_{l,w} + \sqrt{\mathcal{N}}\kappa \delta_{l,l_\mathcal{S}} \nonumber \\
& \hspace{1cm} + \chi_{l_L,w}(t)\delta_{l_L,w} + \chi_{l_R,w}(t)\delta_{l_R,w}, \label{eq:tweveq} 
\end{align}
with $J_{l\pm}$ the function introduced in Eq.\ \eqref{eq:cs} for the implementation of complex scaling. The last line of Eq.\ \eqref{eq:tweveq} describes how the initial vacuum fluctuations outside the scattering region penetrate the system and represent quantum noise that enters the scattering region \cite{Dujardin2015PRA}. It is given by
\begin{align}
\chi_{l_L,w}(t) & = E_\delta e^{-i (2E_\delta-\mu)\tau/\hbar} \sum_{l'=-\infty}^{-1} L_{l'}(\tau) \nonumber \\
& \hspace{1cm} \times \sum_{k= 0}^{W-1} T_k(\tau)\eta_{l',k}(0) e^{2\pi i k w/W} \label{eq:tWNoise1} \\
\chi_{l_R,w}(t) & = -E_\delta e^{-i (2E_\delta-\mu)\tau/\hbar} \sum_{l'=1}^{\infty} L_{l'}(\tau) \nonumber \\
& \hspace{1cm} \times \sum_{k= 0}^{W-1} T_k(\tau)\eta_{l',k}(0) e^{2\pi i k w/W} \label{eq:tWNoise2},
\end{align} 
with $\tau = (t-t_0)$ and
\begin{align}
L_\alpha(t-t_0) & = \dfrac{i^\alpha}{2}\left[J_{\alpha+1}\left(\dfrac{E_\delta\tau}{\hbar}\right) + J_{\alpha-1}\left(\dfrac{E_\delta\tau}{\hbar}\right)\right],
\end{align}
where $J_\nu(\tau)$ are the Bessel functions of the first--kind of order $\nu$ and
\begin{equation}
T_k(t-t_0) = \dfrac{1}{\sqrt{W}}e^{iE_\delta\tau\cos(2\pi k/W)/\hbar}.
\end{equation}
We use classical field amplitudes $\displaystyle \{\psi_{l,w}\}$ that are randomly chosen to properly sample the initial many--body quantum state of the system. At initial time, the scattering region is fully empty and the corresponding Wigner function is a product of vacuum Wigner functions 
\begin{equation}
\mathcal{W}_{SR}(\{\psi_{l,w},\psi_{l,w}^*\},t_0) = \prod_l\prod_m \left(\frac{2}{\pi} e^{-2|\psi_{l,w}|^2}\right).
\label{eq:wsr}
\end{equation}
The source of atoms is populated with a large number $\vert\chi\vert^2=\mathcal{N} \gg 1$ of atoms, which allows one to treat the source as a coherent state whose Wigner function reads
\begin{equation}
\mathcal{W}_\mathcal{S}(\chi,\chi^*,t_0) = \dfrac{2}{\pi} e^{-2|\chi - \sqrt{\mathcal{N}}|^2}.
\label{eq:wsou}
\end{equation}
The Wigner function that describes the whole system is simply given by the product of the Wigner functions \eqref{eq:wsr} and \eqref{eq:wsou}
\begin{align}
\mathcal{W}(\{\psi_{l,w},\psi_{l,w}^*\},t_0) & = \mathcal{W}_{SR}(\{\psi_{l,w},\psi_{l,w}^*\},t_0) \nonumber \\
& \hspace{0.5cm}\times \mathcal{W}_\mathcal{S}(\chi,\chi^*,t_0).
\end{align}  
Consequently, the classical field amplitudes are chosen as
\begin{equation}
\displaystyle \psi_{l,w}(t=t_0) = \dfrac{1}{2}\left(\mathcal{A}_{l,w} + i \mathcal{B}_{l,w}\right),
\end{equation}
where $\mathcal{A}_{l,w}$ and $\mathcal{B}_{l,w}$ are real and independent gaussian random variables fulfilling
\begin{align}
\overline{\mathcal{A}_{l,w}} & = \overline{\mathcal{B}_{l,w}} = 0, \\
\overline{\mathcal{A}_{{l',w'}}\mathcal{A}_{l,w}}  &= \overline{\mathcal{B}_{{l',w'}}\mathcal{B}_{l,w}}  = \delta_{l,l'}\delta_{w,w'}, \\
\overline{\mathcal{A}_{{l',w'}}\mathcal{B}_{l,w}} &= 0,
\end{align}
where $\overline{\cdot}$ denotes an average over the random variables. That choice for the classical field amplitudes implies that a fictious average vacuum population $\overline{\vert\psi_{l,w}(t_0)\vert^2} = 1/2$ is artificially introduced at the initial time. The computation of the atomic density must therefore include a subtraction of this half fictitious particle per site. 

Owing to the large number of atoms that populate the source, we can safely consider that the relative uncertainties of both the amplitude and the phase of the source are negligible. This approximation allows us to treat the source classically and to set $\displaystyle \chi(t=t_0) = \sqrt{\mathcal{N}}$. We additionally choose $\kappa(t)\rightarrow 0$ while keeping $\mathcal{N}\vert\kappa\vert^2$ finite and constant, allowing us to neglect the source depletion and any back--action of the scattering region on the source \cite{Dujardin2015PRA}. In this limit, one can solely focus on the evolution within the scattering region, and the propagation equation for the amplitude of the classical fields on each point of our lattice is therefore given by Eq.\ \eqref{eq:tweveq}.

An average performed over the sampling of the initial many--body quantum state gives access to the observables of interest. We demonstrate this for the $(k,n)$ mode density in the momentum space which is evaluated in a slab of $\tilde{L} \times W$ sites in the upstream region. This mode density is yielded as
\begin{equation}
\tilde{n}_{k,n} = \dfrac{1}{\tilde{L}W}\overline{\left\vert\sum_{l}\sum_{w} \psi_{l,w} e^{-2\pi i(kl/\tilde{L} + nw/W)} \right\vert^2} -\dfrac{1}{2}, 
\label{eq:twdtot}
\end{equation}
where the subtraction of $1/2$ compensates for the artificial $1/2$ atom per site in the momentum space, as explained above. The truncated Wigner method allows one, contrarily to a mean--field approach, to access both coherent and incoherent quantities. The coherent contributions to the $(k,n)$ mode density in momentum space is given by
\begin{equation}
\tilde{n}_{k,n}^{\text{coh}} = \dfrac{1}{\tilde{L}W}\left\vert\overline{\sum_{l}\sum_{w} \psi_{l,w} e^{-2\pi i(kl/\tilde{L} + nw/W)}} \right\vert^2, 
\label{eq:twdcoh}
\end{equation}
and the incoherent one is then obtained through
\begin{equation}
\tilde{n}_{k,n}^{\text{incoh}} = \tilde{n}_{k,n} - \tilde{n}_{k,n}^{\text{coh}}.
\label{eq:twdincoh}
\end{equation}
This notion of coherence is meaningful for matter waves and characterizes the capacity of the atom laser to produce superposition and interference effects. Note that the interaction--induced loss of this matter--wave coherence must not be confused with environment--induced decoherence in the many--body Fock space that would arise if the system is coupled to a heat bath.

\section{Results \label{sec:res}}
\subsection{Coherent backscattering peak}
\begin{figure}[h!]
\begin{center}
\includegraphics[width=9cm]{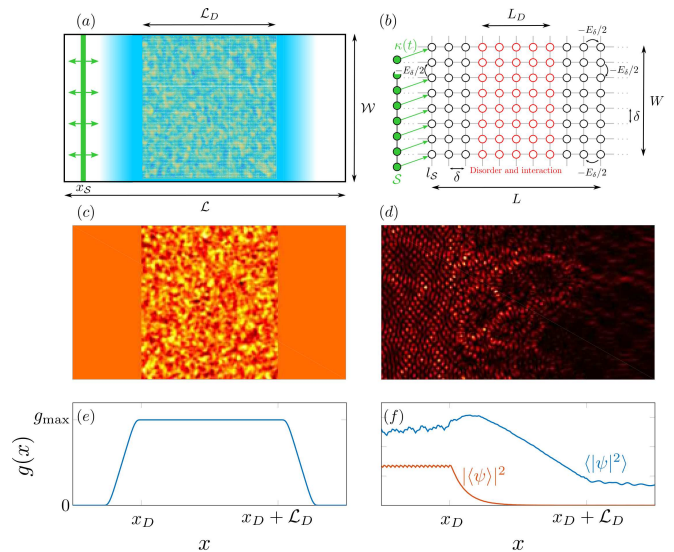}
\end{center}
\caption{Numerical representation of the 2D scattering configuration. A coherent source of bosons is coupled to the scattering region at position $x_\mathcal{S}$ and injects a monochromatic plane wave beam which travels towards a two--dimensional region of space in which disorder and interaction are present, as is depicted in panel (a). The discretisation of the 2D scattering region of length $\mathcal{L}$ and width $\mathcal{W}$ results in a lattice of $L$ (resp. $W$) sites in the longitudinal (resp. transverse) direction with the spacing $\delta$ and a nearest neighbour hopping term $-E_{\delta}/2$ in both directions. Smooth exterior complex scaling is applied in the longitudinal direction for absorbing outgoing waves while periodic boundary conditions are imposed in the transverse direction. Panel (c) shows the scattering geometry consisting of two regions with $V(x,y) = 0$ surrounding a slab of length $\mathcal{L}_D$ and width $\mathcal{W}$ where a smooth random disorder is generated. Panel (e) shows the spatial variation of $g(x)$ that is smoothly ramped from $0$ before the disordered region to $g_\text{max}$ in the disorder and smoothly ramped back to $0$ behind this region. Panel (d) shows a single realisation of the steady scattering state achieved in the presence of the disorder displayed in (c). Panel (f) shows the exponential decay of the coherent mode $\vert\langle \psi_{l,w} \rangle\vert^2$ and the linear decrease of the density $\langle\vert \psi_{l,w} \vert^2\rangle$. Numerical parameters : $k\delta=1$, $\sqrt{\mathcal{N}}\vert\kappa\vert^2 m/\hbar^2 k^2=1$, 1000 realisations of a gaussian correlated disorder with disorder strength $V_0 m /\hbar^2k^2 = 0.1$ and correlation length $k\sigma = 1$, length $k \mathcal{L}_D=100$ and width $k\mathcal{W}=120$.}
\label{fig:dens}
\end{figure}

We first perform a mean--field study. Considering that initially the scattering region, depicted in panel (a) of \textsc{Fig}.\ \ref{fig:dens}, is totally empty, \emph{i.e.} $\psi_{l,w} = 0$ at $t=t_0$, we numerically integrate Eq.\ \eqref{eq:SchrSourterm} on the grid depicted in panel (b) of \textsc{Fig}.\ \ref{fig:dens} for various disorder potentials and perform the disorder averages of the observables under study. The scattering geometry can be represented by a region of space where we consider the presence of a smooth gaussian correlated disorder as is described in Eq.\ \eqref{eq:disorder} surrounded by two regions where $V(x,y) = 0$, as depicted in panel (c) of \textsc{Fig}.\ \ref{fig:dens}. We also consider the presence of an effective interaction strength that is constant and equal to $g_\text{max}$ in the disordered region and that is adiabatically ramped from zero to $g_\text{max}$ upstream from the disordered slab and from $g_\text{max}$ to zero downstream, following the profile (e) of \textsc{Fig}.\ \ref{fig:dens}.

Provided the nonlinearity remains sufficiently small in the Gross--Pitaevskii equation \eqref{eq:SchrSourterm}, there exists a steady stable scattering state \cite{Johansson_2009}. At higher interaction strengths however, dynamical instabilities can occur \cite{SkipetrovPRL2000,Paul2005PRA}, thus rendering a steady scattering state unreachable because the scattering process remains always time--dependent. Since we want to focus on quasi--steady scattering processes, we have to restrict the interaction strength to very low values.

In the absence of nonlinearity, we can, for each disorder realisation, reach a steady scattering state, one of which being displayed in panel (d) of \textsc{Fig}.\ \ref{fig:dens}. Taking the disorder average of these states leads to the coherent mode $\vert\langle \psi_{l,w} \rangle\vert^2$ and to the mean density $\langle\vert \psi_{l,w} \vert^2\rangle$, depending on whether the disorder average is performed before or after the square modulus. The lower right panel of \textsc{Fig}.\ \ref{fig:dens} shows an average over the $y$ direction of $\vert\langle \psi_{l,w} \rangle\vert^2$ and $\langle\vert \psi_{l,w} \vert^2\rangle$. We observe, as was also found in \cite{Hartung2008PRL}, an exponential decay of the coherent mode $\vert\langle \psi_{l,w} \rangle\vert^2 \propto \exp(-x/l_s)$ where $l_s$ is the scattering mean free path. From panel (f) of \textsc{Fig}.\ \ref{fig:dens}, we extract $kl_s \approx 11$, indicating that we are in the so--called $kl_s \gg 1$ weak disorder regime, as well as in the $l_s \ll \mathcal{L}_D$ diffusive regime, which is also confirmed by the linear decrease over the longitudinal direction of the density. This allows us to compute the Boltzmann mean free path which is defined as \cite{akkermans_montambaux_2007}
\begin{equation}
\dfrac{l_s}{l_B} = 1 - \dfrac{I_1(2k^2\sigma^2)}{I_0(2k^2\sigma^2)},
\end{equation}
where $I_\nu(z)$ is the modified Bessel function of order $\nu$, yielding $kl_B \approx 37$. We also extract the transport mean free path $l_{tr}$ \cite{FreundPRL88} using the scaling $\langle\vert \psi_{l,w} \vert^2\rangle \propto \mathcal{L}_D + 0.82 l_{tr} - x$ of the disorder--averaged density and find $kl_{tr} \approx 39$, indicating that the chosen correlation length yields anisotropic scattering. Finally, the localisation length is provided by \cite{Kuhn_2007} $\xi_\text{loc} = l_B \exp(\pi kl_B/2)$ and exceeds, by far, the dimension of the scattering region.

The two--dimensional Fourier transform of the wavefunction is taken in an upstream region where both disorder and nonlinearity are equal to zero. The different Fourier modes can hence be associated to outgoing waves in various directions with the wavenumbers $\mathbf{k}_n =
\sqrt{k^2 - (2\pi n/\mathcal{W})^2}\mathbf{e}_x +(2\pi n/\mathcal{W}) \mathbf{e}_y$, describing the propagation in a spatial direction characterised by the angle $\theta_n = \arcsin[2\pi n /(k\mathcal{W})]$, with $n=-W/2, -W/2+1, \ldots,W/2$.
\begin{figure}[h!]
\begin{center}
\includegraphics[width=6cm]{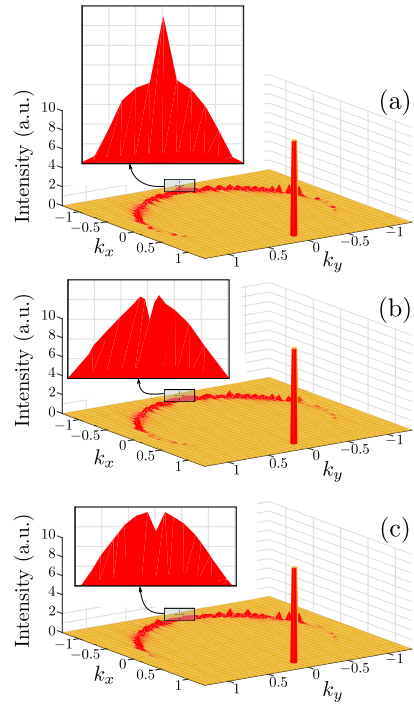}
\end{center}
\caption{Disorder average of the two--dimensional Fourier transform of the quasi--stationary scattering wavefunction evaluated in the upstream region. A strong peak (whose heigth is cut) at $k_x/k = 1$ appears as a clear signature of the incoming plane wave. States forming the circle $k_x^2+k_y^2=k^2$ around the origin are also populated, with a higher value in the backscattered direction. The inset shows a zoom around $(k_x/k,k_y/k) = (-1,0)$ which is the backscattered mode. Panel (a) shows that in the absence of interaction, a peak in the mode $(k_x/k,k_y/k) = (-1,0)$ associated to coherent backscattering appears. In the presence of a small interaction strength $g = 0.005$, Gross--Pitaevskii simulations shown in panel (b) indicate that the coherent backscattering peak is inverted. Truncated Wigner simulations (c) show that this effect is partially destroyed due to many--body interaction effects that are responsible for dephasing of interfering trajectories. Numerical parameters : $k\delta=1$, $\sqrt{\mathcal{N}}\vert\kappa\vert^2 m/\hbar^2k^2=1$, 1500 realisations of a gaussian correlated disorder with disorder strength $V_0 m /\hbar^2k^2 = 0.1$ and correlation length $k\sigma = 1$, length $k\mathcal{L}_D=100$ and width $k\mathcal{W}=120$. In panel (c), we have the injected density $\rho^\varnothing/k^2 \approx 1.33$ and $g = 0.005$.}
\label{fig:CBS2DFFT}
\end{figure}

In the absence of interaction, Gross--Pitaevskii simulations show the appearance of coherent backscattering. This is clearly visible in the panel (a) of \textsc{Fig}.\ \ref{fig:CBS2DFFT}, where we observe that the modes associated to outgoing waves display similar populations, forming a ridge along the circle $k_x^2+k_y^2=k^2$, which indicates that all directions of reflection are approximately equivalently populated. The inset of panel (a) of \textsc{Fig}.\ \ref{fig:CBS2DFFT} shows a zoom around $(k_x/k,k_y/k) = (-1,0)$, corresponding to the backscattered direction, which highlights a higher population of the mode associated with coherent backscattering, as compared to other scattering directions. Artificial oscillations, which are the result of the periodic boundary conditions, are present for large angles, indicating that a more suitable method to extract and analyse coherent backscattering is required.

The heigth of the CBS peak in \textsc{Fig}.\ \ref{fig:CBS2DFFT}(a) is reduced compared to the semiclassical expectation of a factor 2 enhancement. This reduction is due to the presence of short length self--retracing paths, mainly those that feature a backreflection at only a single scattering event within the disordered region. Those paths are identical to their time--reversed counterpart and bring no contribution to CBS. Their relative weight in the sum over all backreflected paths gives therefore rise to a reduction of the CBS enhancement.

It has been argued in Ref.\ \cite{Hartmann2012AP} that this reduction of the CBS peak height, induced by self--retracing paths, should be quantitatively identical to the depth of the dip that forms in the presence of mean--field interaction. The occurrence of this dip is shown in \textsc{Fig}.\ \ref{fig:CBS2DFFT}(b) which displays that the CBS peak becomes a pronounced dip in the presence of interaction, as was also observed in Refs.\ \cite{Hartung2008PRL}. Beyond the mean--field regime, truncated Wigner simulations depicted in panel (c) indicate that this inversion prevails. It is however partially destroyed due to many--body interaction effects that create incoherent particles, from which results dephasing.

Despite the fact that the Truncated Wigner method accounts for off--shell scattering events between the atoms \cite{Dujardin2015PRA}, which populate states with a kinetic energy different from that of the incident particles, we do not observe a significant broadening of the density distribution about the energy shell in the presence of interaction \cite{Geiger2013NJoP}. We attribute this to the fact that the atoms do not stay long in the disordered region where they interact. This thermal cloud around the condensate is expected to be more pronounced in a turbulent regime which we do not study here. As a matter of fact, the set of parameters chosen in \textsc{Fig}.\ \ref{fig:CBS2DFFT}(c) yields a mostly coherent current, as is confirmed in \textsc{Fig}.\ \ref{fig:tWg0.005}(c).

\subsection{Angular resolved current}
The drawback of the two--dimensional Fourier transform is that it demands a large number of sites to yield a satisfactory resolution. An alternative way to extract the reflected part of the wavefunction is to take the partial Fourier transform $\tilde{\psi}(x,k_y)$ of $\psi(x,y)$ along the y--direction. This new wavefunction contains both the incident part ($+$) and the reflected part ($-$) and one should get rid of the former. Considering $\tilde{\psi}(x,k_y)$ at position $x_0$ and position $x_1 = x_0 + \Delta$, we have, introducing $\alpha_\pm$, the amplitudes of the incident and reflected waves
\begin{equation}
\begin{pmatrix}
\tilde{\psi}(x_1,k_y) \\
\tilde{\psi}(x_0,k_y)
\end{pmatrix}
=
\begin{pmatrix}
e^{ik_x\Delta} & e^{-ik_x\Delta} \\
1 & 1
\end{pmatrix}
\begin{pmatrix}
\alpha_+ \tilde{\psi}^{(+)}(x_0,k_y) \\
\alpha_- \tilde{\psi}^{(-)}(x_0,k_y)
\end{pmatrix},
\end{equation}
whose solution is found to be
\begin{align}
\begin{pmatrix}
\alpha_+ \tilde{\psi}^{(+)}(x_0,k_y) \\
\alpha_- \tilde{\psi}^{(-)}(x_0,k_y)
\end{pmatrix} 
& =
\dfrac{1}{2i\sin(k_x\Delta)}
\begin{pmatrix}
1 & -e^{-ik_x\Delta} \\
-1 & e^{ik_x\Delta}
\end{pmatrix} \nonumber \\
& \hspace{2.5cm}\begin{pmatrix}
\tilde{\psi}(x_1,k_y) \\
\tilde{\psi}(x_0,k_y)
\end{pmatrix}.
\end{align}
This allows us to separate the incoming and reflected components of the wavefunction at position $x_0$, the latter being given by
\begin{equation}
\alpha_- \tilde{\psi}^{(-)}(x_0,k_y) = \dfrac{\tilde{\psi}(x_0,k_y)e^{ik_x\Delta} - \tilde{\psi}(x_1,k_y)}{2i\sin(k_x\Delta)}.
\end{equation}
The current density $j_n$ in the direction $\theta_n$ is given by
\begin{equation}
j_n =  2\pi\dfrac{\hbar}{m} \sqrt{k^2-k_y^2} \vert \tilde{\psi}_n \vert^2 \cos\theta_n,
\label{eq:foumod}
\end{equation}
where we have written $\tilde{\psi}_n \equiv \alpha_- \tilde{\psi}^{-}(x_0,k_y)$ and $\theta_n = \arcsin(k_y/k)$. We finally note that $x_0$ and $x_1$ must be chosen in a region where the interaction (and consequently the disorder) is equal to zero. This allows us to apply the superposition principle and to associate the Fourier modes in Eq.\ \eqref{eq:foumod} to directions in the two--dimensional space. In the following, we choose $kx_0 = 10$ and $kx_1 = 20$.

\begin{figure}[h!]
\includegraphics[width=8.5cm]{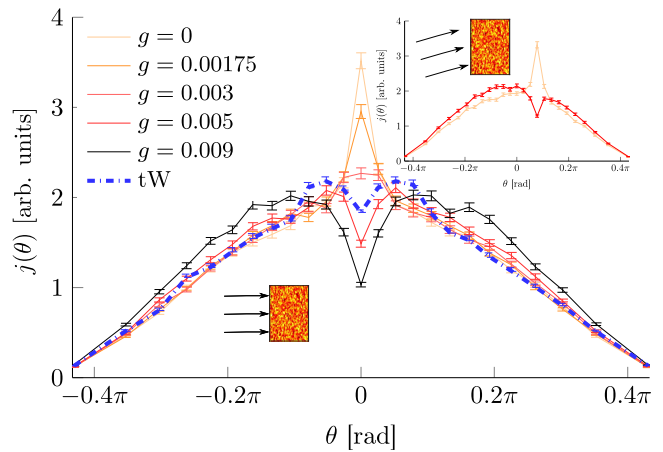}
\caption{Angular resolved current as a function of the backscattered angle $\theta_n = \arcsin[2\pi n /(k \mathcal{W})]$ for different values of the interaction strength $g$. Error bars indicate the statistical standard deviation. In the absence of nonlinearity, we recover the typical coherent backscattering cone at $\theta = 0$. For increasing values of the nonlinearity, we find, reproducing the behaviour observed in Ref. \cite{Hartung2008PRL}, that the peak is first damped and then inverted into a dip, indicating that constructive interferences become destructive. We also show, in dash--dotted line, a truncated Wigner curve that shows the total backscattered current, which, despite a partial dephasing, is in quite good agreement with its Gross--Pitaevskii counterpart. We show in the inset that this effect is indeed due to interference between reflected paths and not due to the geometry of the slab : if we tilt the source by the angle $\phi_{-3} = \arcsin[ 2\pi(-3)/(k \mathcal{W})] \approx -0.16$, we get the peak and the dip at the exact opposite angle. Numerical parameters : $k\delta=1$, $\sqrt{\mathcal{N}}\vert\kappa\vert^2 m/\hbar^2 k^2=1$, 1500 realisations of a gaussian correlated disorder with disorder strength $V_0 m /\hbar^2k^2 = 0.1$ and correlation length $k\sigma = 1$, length $k\mathcal{L}_D=40$ and width $k\mathcal{W}=120$. For the truncated Wigner simulation : $g = 0.005, \rho^\varnothing/k^2 \approx 1.33$.}
\label{fig:inv}
\end{figure}
The angular resolved current is depicted in \textsc{Fig.}\ \ref{fig:inv} for different values of the interaction strength $g$. In the noninteracting case, we recover the characteristic coherent backscattering peak we already observed in the inset of \textsc{Fig.}\ \ref{fig:CBS2DFFT} (a). For higher interaction strengths, the peak turns to a dip, as was also observed in Ref. \cite{Hartung2008PRL}, indicating a crossover from constructive to destructive interferences. Considering a tilt of the source by an angle $\phi_{-3} = \arcsin[ 2\pi(-3)/(k\mathcal{W})] \approx -0.16$, which amounts to choosing the tilted profile $\phi(y) = \exp\left[(i2\pi (-3)/k\mathcal{W})y\right]$ in Eq.\ \eqref{eq:idealised}, the inset of \textsc{Fig.}\ \ref{fig:inv} shows that the coherent backscattering peak and the related inversion are realised in the exact opposite direction. This confirms the interference effect between scattering paths and their time--reversed counterparts and validates that coherent backscattering is the underlying mechanism.

\subsection{Inversion of coherent backscattering beyond the mean--field regime}
While it is already explained in Sec.\ \ref{sec:tw}, namely in Eqs. \eqref{eq:twdtot}, \eqref{eq:twdcoh} and \eqref{eq:twdincoh}, how to compute the total, coherent and incoherent $(k, n)$ mode densities in the momentum space, the procedure for doing so for the current requires further explanations. Depending on whether the average is performed over the wavefunctions (the square modulus being taken on the average wavefunction) or over the square modulus of those, one can define total and coherent current, similarly as for the Gross--Pitaevskii simulations
\begin{align}
j_{n}^\text{tot} & = 2\pi\dfrac{\hbar}{m} \sqrt{k^2-k^2_n} \left(\overline{\vert\tilde{\psi}_n\vert^2} - \dfrac{1}{2}\right) \cos \theta_n \\
j_{n}^\text{coh} & = 2\pi\dfrac{\hbar}{m} \sqrt{k^2-k^2_n}\vert\overline{\tilde{\psi}_n}\vert^2  \cos \theta_n.
\end{align}
The incoherent part of the current is obtained by subtracting the coherent contribution from the total one
\begin{equation}
j_n^\text{incoh} = j_n^\text{tot} - j_n^\text{coh}.
\end{equation}
This allows us to investigate to which extent the inverted structure is due to coherent contribution, and to find out which interaction strength leads to a dephasing between interfering trajectories and finally yields a structureless current, with a dominant incoherent contribution, as was predicted by a nonlinear diagrammatic theory in \cite{Geiger2013NJoP} and numerically confirmed in \cite{ChretienPRA18} for Al'tshuler--Aronov--Spivak oscillations (see also Ref.\ \cite{ScottPRA2008} in this context). The identification of this dephasing regime is fundamental, as it provides information whether the effect is experimentally observable.

\begin{figure}[h!]
\begin{center}
\includegraphics[width=0.9\linewidth, clip=true]{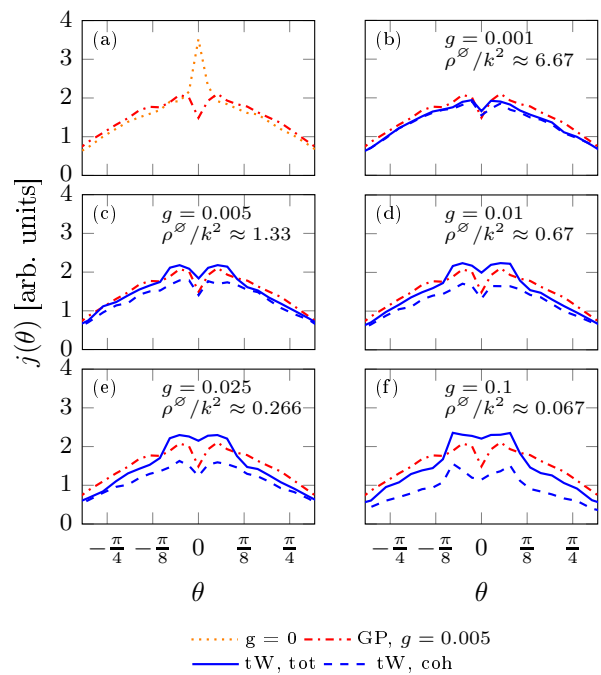}
\end{center} 
\caption{Truncated Wigner simulations of the angular--resolved backscattered current for an increasing interaction strength $g$ and a correspondingly decreasing average density $\rho^\varnothing$, the product $g\rho^\varnothing/k^2 \simeq 0.00665$ being kept constant for all simulations. The dotted orange and dash--dotted red curves show Gross--Pitaevskii simulations for $g = 0$ and $g = 0.005$ in panel (a), that is, in a regime where backscattering is first observed and then partially inverted. The dip appearing in panel (a) in the presence of interaction is preserved beyond the mean--field regime, but is partially destroyed due to dephasing resulting from an increasing incoherent contribution. Numerical parameters : $k\delta=1$, 500 realisations of a gaussian correlated disorder with disorder strength $V_0 m /\hbar^2k^2 = 0.1$ and correlation length $k\sigma = 1$, length $k\mathcal{L}_D=40$ and width $k\mathcal{W}=120$.}
\label{fig:tWg0.005}
\end{figure}

We first perform truncated Wigner simulations for the case of a partial inversion of coherent backscattering that corresponds to the red (dark grey) curve of \textsc{Fig}.\ \ref{fig:inv}, that is, a scenario with $\rho^\varnothing g/k^2 \simeq 0.00665$. We vary both the density per unit surface $\rho^\varnothing$ and the interaction strength $g$ while maintaining the nonlinearity $\rho^\varnothing g/k^2$ constant, which allows us to explore the many--body effects beyond the mean--field regime. \textsc{Fig}.\ \ref{fig:tWg0.005} shows that a partial inversion of coherent backscattering prevails beyond the mean--field regime. Panel (f) of \textsc{Fig}.\ \ref{fig:tWg0.005} indeed indicates that even for densities as low as $\rho^\varnothing/k^2 \simeq 0.067$, the inversion is still preserved, although a certain dephasing has already appeared, thereby partially destroying the effect. 

We now evaluate whether this effect is observable with $^{87}$Rb. Considering the 2D interaction strength which we can write as
\begin{equation}
\tilde{g}(x) = \dfrac{\hbar^2}{m} g(x) = \dfrac{\hbar^2}{m} \dfrac{2\sqrt{2\pi}a_S}{a_\perp(x)},
\end{equation}
with $a_\perp(x) = \sqrt{\hbar/m\omega_\perp(x)}$ the oscillator length associated to the confinement frequency $\omega_\perp(x)$ of the trap, we can write within the disordered region
\begin{equation}
\tilde{g}(x)\rho^\varnothing/k^2 = \dfrac{\hbar^2}{m} 2\sqrt{2\pi} \dfrac{a_S}{a_\perp(x)}\rho^\varnothing/k^2.
\end{equation}
In the simulations, we have the chemical potential $\mu = E_\delta/2 =  mv^2/2$, corresponding to $k = mv/\hbar$ with $v$ the velocity of the injected particles. In the case we investigate here, where $\rho^\varnothing g(x)/k^2= 0.00665$, the injected density is found to be
\begin{equation}
\rho^\varnothing/k^2 = \dfrac{0.00665}{2\sqrt{2\pi}} \dfrac{a_\perp}{a_S},
\end{equation}
and essentially depends on the s--wave scattering length and the oscillator length which scales as $\omega_\perp^{-1/2}$. Considering the s--wave scattering length $a_S = 5.313 \times 10^{-9}$ m of $^{87}$Rb and its mass $m = 1.443 \times 10^{-25}$ kg, we find that for a confinement frequency of $\omega_\perp/2\pi = 75$ Hz, the injected density reaches $\rho^\varnothing/k^2 \approx 0.31$. This corresponds to a situation similar as that depicted in the panel (e) of \textsc{Fig}.\ \ref{fig:tWg0.005}. We therefore believe that such an inversion of coherent backscattering should be observable experimentally.

\begin{figure}[h!]
\begin{center}
\includegraphics[width=0.9\linewidth, clip=true]{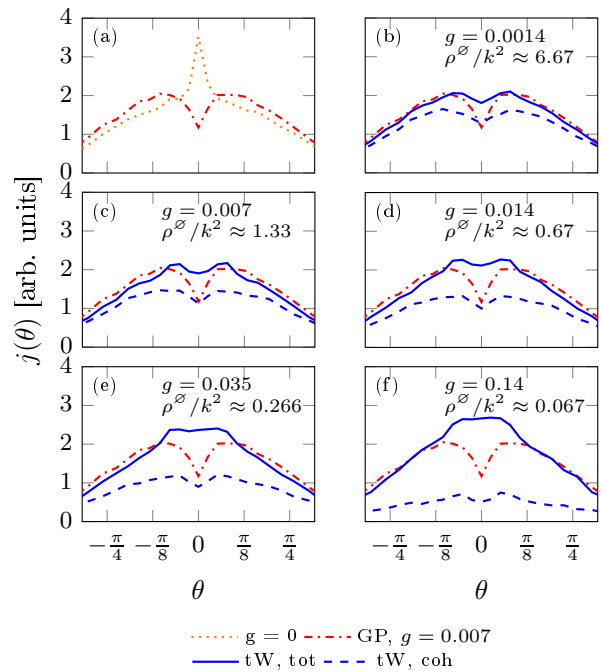}
\end{center} 
\caption{Same as \textsc{Fig}.\ \ref{fig:tWg0.005}, but for $g\rho^\varnothing/k^2 \simeq 0.00931$. While a more pronouced dip is found with the Gross--Pitaevskii red curve, more dephasing is observed in the results of the truncated Wigner simulations.}
\label{fig:tWg0.007}
\end{figure}

We also note that we have to enforce $\frac{1}{2} m v^2 < \hbar \omega_\perp$ in order to safely neglect the excitation of the transverse modes of the condensates. With the parameters we used, the choice of a velocity for the injected particles of $v = \sqrt{\hbar\omega_\perp/m} = 0.00068$ m/s satisfies this constraint. We should note that one would still be in a supersonic regime with such a velocity. Indeed, the speed of the sound within the condensate is given by $v_c = \left(\rho^\varnothing g_{3D}/m\sqrt{\pi}a_\perp\right)^{1/2}$, where the $1/\sqrt{\pi}a_\perp$ factor comes from the transverse wavefunction in its ground state $\phi(z) = e^{-z^2/2a_\perp^2}/\sqrt{\sqrt{\pi}a_\perp}$ evaluated for $z = 0$ and where $g_{3D} = 4\pi\hbar^2a_S/m$ is the 3D interaction strength. This allows us to rewrite the speed of sound within the condensate as $v_c = (\sqrt{2} g \rho^\varnothing / k^2)^{1/2} v$, which yields $v_c \simeq 0.1 v$ for parameters used in \textsc{Fig}.\ \ref{fig:tWg0.005} and \textsc{Fig}.\ \ref{fig:tWg0.007}. This confirms that an inversion of coherent backscattering might be observed with $^{87}$Rb. 

\textsc{Fig}.\ \ref{fig:tWg0.007} is dedicated to a truncated Wigner study of the regime corresponding to a full inversion of coherent backscattering. As could be inferred from \textsc{Fig}.\ \ref{fig:tWg0.005}, the dephasing regime is reached with higher nonlinearities. As panel (f) indicates, the coherent contribution carrying the inverted dip is now hidden behind a flat and structureless incoherent contribution that overshadows the signature of interference and interaction. Conducting the same reasoning as for \textsc{Fig}.\ \ref{fig:tWg0.005} with a nonlinearity equal to $g\rho^\varnothing/k^2 = 0.00931$ leads to an injected density $\rho^\varnothing/k^2 \approx 0.44$, which corresponds to a situation intermediate between those depicted in the panels (d) and (e) of \textsc{Fig}.\ \ref{fig:tWg0.007}. Pushing the system further in the quantum limit by increasing the interaction strength and decreasing accordingly the injected density induces, however, dephasing, as panel (f) indicates, where the coherent inversion of backscattering is drowned by an incoherent contribution.

\section{Conclusions \label{sec:concl}}
In conclusion, we numerically studied the two--dimensional transport of Bose--Einstein condensates across a disordered region, which gives rise to a weak--localisation scenario that occurs through coherent backscattering. In the mean--field regime, which is studied by means of the Gross--Pitaevskii equation, the presence of an atom--atom interaction gives rise to a crossover around $g\rho^\varnothing/k^2 \simeq 0.004$ from constructive to destructive interference, the peak in the backscattered current becoming a pronounced dip, thereby reverting weak localisation \cite{Hartung2008PRL,Hartmann2012AP}. Truncated Wigner simulations show that the coherent backscattering inversion is also encountered when accounting for effects beyond the mean--field approximation. As was observed in Refs.\ \cite{ChretienPRA18,Geiger2013NJoP} when pushing the limit far beyond the mean--field regime, quantum interference effects face dephasing, and the dip structure is completely overshadowed by a dominant incoherent contribution.

We believe that this effect is experimentally observable for $^{87}$Rb. We indeed found for experimentally realistic parameters a value for the injected density, namely $\rho^\varnothing/k^2 \simeq 0.44,$ for which the inversion is predicted to be still observable. Other species, such as $^{39}$K, whose s--wave scattering length can be tuned by means of Feshbach resonances to very low values, are also good candidates to realise a full inversion of coherent backscattering. 

The present study had a clear focus on the inversion of the CBS peak in the framework of a quasi--stationary 2D propagation of a Bose--Einstein condensate across a disordered region. It thereby left out a number of interesting side investigations that one could have performed in this context with our numerical setup. Among these are the study of wavepacket propagation processes across the disorder potential, in alignement with the experiments of Ref.\ \cite{PhysRevLett.109.195302} (see also \cite{WellensPRA2016}), as well as coherent forward scattering \cite{CherroretPRL2012,CherroretPRA2014,MiniaturaPRA2014,MuellerPRL2014,WellensPRA2016} which is expected to occur in the downstream region behind the disordered slab. A comparison with the study undertaken in Ref.\ \cite{SkipetrovPRL2008}, dedicated to the expansion of a $k$--resolved source where it is argued that as the cloud expands for a sufficiently long time interaction disappears, would also be relevant. Furthermore, this study lacks a quantitative comparison with diagrammatic many--body scattering theory \cite{Geiger2013NJoP}, which is, however, difficult to carry out because of the inhonomegenous density profile of the Bose--Einstein condensate. In that context, it would therefore be interesting to investigate the transport of Bose--Einstein condensates through 2D billiard potentials \cite{Hartmann2012AP} or in the non-equilibrium configuration described in Ref.\ \cite{ScoquartPRR2020}, where a homogeneous density profile is expected.

\begin{acknowledgments}
The computational resources have been provided by the Consortium des Equipements de Calcul Intensif (CÉCI), funded by the F.R.S.-FNRS under Grant No. 2.5020.11. 
\end{acknowledgments}

\appendix

\section{Numerical scheme \label{sec:NumScheme}}
The discretisation scheme of the partial differential equations in terms of finite differences naturally results in a set of ordinary differential equations which we numerically integrate. For that purpose, we use a very general numerical scheme based on the expansion of the solution in a Taylor series \cite{HDMeyer}. This method allows in principle to determine the solution of every ordinary differential equation, but requires on the other hand to compute the derivatives up to the desired order, which may be tedious and inefficient in some cases. Automatic differention techniques \cite{Wengert64,Barton1971,Bartholome2000,Bruecker2006,Naumann2012} may be envisaged to circumvent this major drawback. Knowing the solution $y(t)$ of the differential equation at time $t$, it is obtained at time $t+\delta t$ using the expansion
\begin{equation}
y_{n+1} = y_n + \delta t \dfrac{d y_n}{d t} + \dfrac{1}{2}(\delta t)^2 \frac{d^2 y_n}{d t^2} + \ldots,
\end{equation}
that should be repeated iteratively from $t=t_0$ until reaching $t=t_f$. Following this principle, the discrete wavefunction at site $(l,w)$ is expanded in Taylor series and is therefore written at time $t + \delta t$ as
\begin{align}
\psi_{l,w}(t + \delta t) & = \psi_{l,w}(t) + \delta t \frac{d\psi_{l,w}}{dt}(t) + \dfrac{1}{2}(\delta t)^2 \dfrac{d^2\psi_{l,w}}{dt^2}(t) + \ldots \nonumber \\
& = \sum_{k=0}^{K_\text{max}} \dfrac{1}{k!} (\delta t)^k \frac{d^k \psi_{l,w}}{dt^k}(t) + \mathcal{O}\left[(\delta t)^{K_\text{max} + 1}\right]
\label{eq:taylor}
\end{align}
and will be propagated from initial time $t_0$ to final time $t_f$ by means of this equation. In Eq. \eqref{eq:taylor}, $K_\text{max}$ denotes the maximal order considered for the derivative in the Taylor expansion. This expansion requires that we are able to compute the derivatives of $\psi_{l,w}$ up to order $K_\text{max}$ which is readily achieved by differentiating the field equations \eqref{eq:tweveq} which gives, for instance, for the second time derivative
\begin{align}
i\hbar \ddot{\psi}_{l,w} & = \left(2E_\delta - \mu + V_{l,w}\right) \dot{\psi}_{l,w} \nonumber \\
& \hspace{1cm} - \dfrac{E_\delta}{2}\left(\dot{\psi}_{l+1,w} + \dot{\psi}_{l-1,w}\right) \nonumber \\
& \hspace{1cm} - \dfrac{E_\delta}{2}\left(\dot{\psi}_{l,w+1} + \dot{\psi}_{l,w-1}\right) \nonumber \\
& \hspace{1cm} + g_{l} \dfrac{d}{dt}(\vert\psi_{l,w}\vert^2\psi_{l,w}) \nonumber \\
& \hspace{1cm} + \dot{\chi}_{l_l,w}(t)\delta_{l_l,w} + \dot{\chi}_{l_R,w}(t)\delta_{l_R,w},
\label{eq:gpwithsecs2d}
\end{align}
where $\dot{}$ denotes the derivatives with respect to $t$ and where we have omitted the derivative of the source term $ \sqrt{\mathcal{N}}\kappa(t) \delta_{l,l_\mathcal{S}}$ because we assume that the coupling $\kappa(t)$ varies so slowly that its derivative with respect to $t$ is negligible. 

While derivatives of the wavefunction $\psi_{l,w}$ are easily found iteratively, the derivatives of the nonlinear term as well as that of the noise terms are more complicated to obtain. Exploiting the property
\begin{align}
\frac{d^n}{dt^n} A(t)B(t)C(t) & = \sum_{k=0}^n \sum_{\lambda = 0}^{n-k} \binom{n}{k,\lambda,n-k-\lambda} \dfrac{d^k}{dt^k} A(t) \nonumber \\
& \hspace{1cm} \times \dfrac{d^\lambda}{dt^\lambda}B(t) \dfrac{d^{n-k-\lambda}}{dt^{n-k-\lambda}}C(t),
\end{align} 
 where $\binom{n}{k,\lambda,n-k-\lambda}$ is the trinomial coefficient, the $k^\text{th}$ derivative of the nonlinear term reads
\begin{align}
\dfrac{d^n}{dt^n} \psi_{l,w}^*(t)\psi_{l,w}^2(t) & = \sum_{k=0}^n \sum_{\lambda = 0}^{n-k} \binom{n}{k,\lambda,n-k-\lambda} \dfrac{d^k}{dt^k} \psi_{l,w}^*(t) \nonumber \\ 
& \hspace{1cm} \times \dfrac{d^\lambda}{dt^\lambda}\psi_{l,w}(t) \dfrac{d^{n-k-\lambda}}{dt^{n-k-\lambda}}\psi_{l,w}(t).
\end{align}
One also has to compute the $k^\text{th}$ derivative of the noise terms in Eqs. \eqref{eq:tWNoise1} and \eqref{eq:tWNoise2}. Noting that the writing of those terms suggests that an inverse discrete Fourier transform has been performed, one can compute the time derivative of the Fourier coefficients of the noise term
\begin{align}
\hat{\chi}_{l_L,k}(t) & = E_\delta e^{-\frac{i}{\hbar}(2E_\delta-\mu)(t-t_0)} T_k(t-t_0) \nonumber \\
& \hspace{2cm} \times \sum_{l'=-\infty}^{-1} L_{l'}(t-t_0)\eta_{l',k}(t_0) \nonumber \\
\hat{\chi}_{l_R,k}(t) & = -E_\delta e^{-\frac{i}{\hbar}(2E_\delta-\mu)(t-t_0)} T_k(t-t_0) \nonumber \\
& \hspace{2cm} \times \sum_{l'=1}^{\infty} L_{l'}(t-t_0)\eta_{l',k}(t_0).
\label{eq:chi5}
\end{align}
which turn out to be
\begin{align}
\dfrac{d^n}{dt^n} \hat{\chi}_{l_L,k}(t) & = E_\delta \left[\sum_{l'=-\infty}^{-1}\left(\sum_{k=0}^n\sum_{\lambda=0}^{n-k}\binom{n}{k,\lambda,n-k-\lambda} \right.\right. \nonumber \\
& \hspace{1cm} \times \dfrac{d^k}{dt^k}L_{l'}(t-t_0)\dfrac{d^\lambda}{dt^\lambda}T_{k}(t-t_0)  \nonumber \\
& \hspace{1cm} \times \left.\left. \dfrac{d^{n-k-\lambda}}{dt^{n-k-\lambda}}e^{-i(2E_\delta-\mu)t/\hbar}\right)\eta_{l',k}(t_0)\right] \label{eq:derker1} \\
\dfrac{d^n}{dt^n} \hat{\chi}_{l_R,k}(t) & = -E_\delta \left[\sum_{l'=1}^{\infty}\left(\sum_{k=0}^n\sum_{\lambda=0}^{n-k}\binom{n}{k,\lambda,n-k-\lambda} \right.\right. \nonumber \\
& \hspace{1cm} \times \dfrac{d^k}{dt^k}L_{l'}(t-t_0)\dfrac{d^\lambda}{dt^\lambda}T_{k}(t-t_0)  \nonumber \\
& \hspace{1cm} \times \left.\left. \dfrac{d^{n-k-\lambda}}{dt^{n-k-\lambda}}e^{-i(2E_\delta-\mu)t/\hbar}\right)\eta_{l',k}(t_0) \right] \label{eq:derker2}
\end{align}
once again involving $\binom{n}{k,\lambda,n-k-\lambda}$, the trinomial coefficient, as well as time derivatives of $L_{l'}$ in the longitudinal direction and $T_k$ in the transverse direction. In Eq.\ \eqref{eq:derker1} and \eqref{eq:derker2}, $\eta_{l',j'}(t_0)$ denotes the initial condition for the wavefunction, which in the mean--field approximation, is identically equal to zero, thereby yielding $\chi_{l_l,w}(t) = \chi_{l_R,m}(t) = 0$, for all $t \geq t_0$. In the truncated Wigner context, however, we have seen that those classical fields are sampled as prescribed in Sec.\ \ref{sec:tw}, with a different sampling from one realisation of the initial condition to another, the convolution kernel remaining unchanged.

\section{A smooth switching function \label{app:timo}}
\begin{figure}[h!]
\begin{center}
\includegraphics[width=0.9\linewidth]{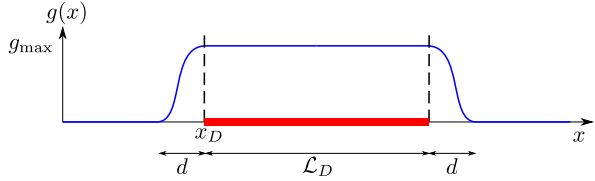}
\end{center}
\caption{Graphical representation of the function $F(x)$.}
\label{fig:smooth}
\end{figure}

The purpose of this Appendix is to present the smooth switching function that we use in this work, which was derived by Hartmann \cite{HartmannThesis}. Considering an interval $I \subset \mathbb{R}$, we are looking for a function $F\in \mathcal{C}^\infty(\mathbb{R})$ that is exactly zero farther than $d>0$ from the interval and that smoothly reaches $1$ over that distance within the interval, as is shown in \textsc{Fig}.\ \ref{fig:smooth}. 

We start by defining the auxiliary test function
\begin{equation}
t(x) = 
\begin{cases}
\exp \left[ -b \left( ax^2 + \dfrac{1}{1-x^2} \right)\right] & \text{for } \vert x \vert < 1 \\
0 & \text{for } \vert x \vert \geq 1 
\end{cases},
\end{equation}
where $a = 0.557747$ and $b = 1.364054$ are numerical parameters that are chosen so that $t(x)$ is as smooth as possible. Based on this test function, we build the function $f:[0,1] \rightarrow \mathbb{R}$ which is defined as
\begin{equation}
f(x) = \dfrac{t(x)}{t(x-1)+t(x)}.
\end{equation}
The smooth switching function describing describing the spatial profile of the interaction is then defined as
\begin{equation}
g(x) = g_\text{max}
\begin{cases}
0 & \text{for } x \leq x_D - d  \\
f\left( \frac{x_D - x}{d}\right) & \text{for } x_D - d < x \leq x_D \\
1 & \text{for } x_D < x \leq x_D + \mathcal{L}_D \\
f\left( \frac{x - x_D + \mathcal{L}_D}{2\mathcal{L}_D+d}\right) & \text{for } x_D < x-\mathcal{L}_D \leq x_D + d \\
0 & \text{for } x_D + \mathcal{L}_D + d < x
\label{eq:func1}
\end{cases}.
\end{equation}
The smooth switching function describing the ramping of the source is given by
\begin{equation}
\kappa(t) = \kappa_\text{max}
\begin{cases}
0 & \text{for } t \leq 0  \\
f\left( \dfrac{t_s-t}{t_s}\right) & \text{for } 0 \leq t \leq t_s \\
1 & \text{for } t_s \leq t
\label{eq:func2}
\end{cases}.
\end{equation}
Specifically, we choose $t_s\mu/\hbar = 400$ for a smooth ramping and typically obtain a stationary scattering state after $t_f\mu/\hbar = 900$.

%
%

\section{Effective atom--atom interaction strength on the square lattice} \label{app:renorm}


In this appendix we discuss how to properly choose the interaction parameter
on the numerical grid that we introduced in order to implement the Truncated
Wigner method.
As each grid point covers a square of the area $\delta^2$, with $\delta$
the lattice spacing, it appears most natural to define the on--site interaction
parameter as
\begin{equation}
  U = \tilde{g} / \delta^2 = g E_\delta =
  \frac{4\pi \hbar^2 a_s}{\sqrt{2\pi} m \delta^2 a_\perp} \,,
  \label{eq:Unaive}
\end{equation}
where $\tilde{g}$ is the effective
two-dimensional atom-atom interaction strength, assuming that the atomic
cloud is located in the transverse ground state of the 2D confinement,
and $g$ denotes its dimensionless counterpart defined according to
Eq.~\eqref{eq:g2D}.
This na\"ive choice, which is retained within Eqs. \eqref{eq:Ham}, \eqref{eq:evop1}, \eqref{eq:GP1} and \eqref{eq:tweveq} for the sake of
simplicity, is problematic insofar as it exhibits convergence issues in the
formal limit $\delta \to 0$, as was discussed in detail in Ref.~\cite{Ron017JPB}.

To determine the correct scaling of the on--site interaction parameter
$U$ with the grid spacing, it is useful to study a two--body scattering
problem on the chosen numerical grid and compare the outcome of this
study with the solution of this problem in the continuous 2D space,
which was obtained in Refs.~\cite{PetHolShl00PRL,PetShl01PRA}.
Considering a harmonic confinement potential with the frequency $\omega_\perp$
in the transverse direction, the Hamiltonian describing this two--body system
in the full three--dimensional space reads
\begin{equation}
  \hat{\mathcal{H}} = - \frac{\hbar^2}{2 m}\left( 
  \frac{\partial^2}{\partial \vec{r}_1^2}
  + \frac{\partial^2}{\partial \vec{r}_2^2} \right) 
  + \frac{1}{2} m \omega_\perp^2 \left( z_1^2 + z_2^2 \right)
  + U(\vec{r}_1 - \vec{r}_2) \label{eq:H}
\end{equation}
with $\vec{r}_j = (x_j,y_j,z_j) \equiv (\mathbf{r}_j,z_j)$ the position 
coordinates of the atom no.~$j=1,2$ and $U$ the two--body interaction strength.
As the latter depends only on the distance between the two atoms, it is
useful to separate the center--of--mass and relative coordinates of the two
atoms and thereby map the collision process between the two atoms into an 
effective one--body scattering problem in the relative coordinates.
The exact solution of the latter in continuous space can then be compared
with the analytic solution of the analogous scattering problem in the
presence of a square--lattice discretization of the in--plane relative 
coordinates.

We should keep in mind, however, that the latter is not exactly equivalent
to the original discretization procedure carried out in the individual 
in--plane coordinates $\mathbf{r}_1$ and $\mathbf{r}_2$ of the atoms, which 
is effectively employed for our numerical simulations.
Indeed, the conventional transformation
$(\vec{r}_1,\vec{r}_2) \mapsto (\vec{R},\vec{r})$ to the 
center--of--mass coordinates $\vec{R} = (\vec{r}_1+\vec{r}_2)/2$ and 
relative coordinates $\vec{r} = \vec{r}_1 - \vec{r}_2$ maps squares into
non--equilateral rectangles and therefore does not preserve the spacing scales
of the discretization procedure.
An approximate equivalence of the two square--lattice discretizations can
nevertheless be established by redefining the new coordinates $\vec{R}$
and $\vec{r}$ in a more symmetric manner, namely through
\begin{eqnarray}
  \vec{R} & = & \frac{1}{\sqrt{2}}(\vec{r}_1+\vec{r}_2) 
  \equiv (\mathbf{R},Z) \,, \label{eq:Rs} \\
  \vec{r} & = & \frac{1}{\sqrt{2}}(\vec{r}_1-\vec{r}_2) 
  \equiv (\mathbf{r},z)\,,\label{eq:rs}
\end{eqnarray}
which corresponds to a unitary mapping that preserves the shape of the
lattice squares resulting from the discretization.
We therefore adopt this latter symmetric definition of 
``center--of--mass'' and ``relative'' coordinates in the following.
The Hamiltonian \eqref{eq:H} can then be separated as 
$\hat{\mathcal{H}} = \hat{H}_c + \hat{H}_r$ with the Hamiltonians
\begin{eqnarray}
  \hat{H}_c & = & - \frac{\hbar^2}{2 m} 
  \frac{\partial^2}{\partial \vec{R}^2} + \frac{1}{2} m \omega_\perp^2 Z^2 
  \label{eq:Hc} \,, \\
  \hat{H}_r & = & - \frac{\hbar^2}{2 m} 
  \frac{\partial^2}{\partial \vec{r}^2} + \frac{1}{2} m \omega_\perp^2 z^2 
  + U(\sqrt{2}\vec{r}) \label{eq:Hr}
\end{eqnarray}
that govern the dynamics in the center--of--mass and relative coordinates,
respectively.

The interaction is modeled via the Fermi--Huang pseudopotential
\begin{equation}
  U(\vec{\rho}) = g_{3D} \delta(\vec{\rho}) \frac{\partial}{\partial \rho} \rho
  \label{eq:FH}
\end{equation}
with
\begin{equation}
  g_{3D} = \frac{4\pi \hbar^2 a_s}{m} \label{eq:g}
\end{equation}
where $a_s$ is the $s$--wave scattering length of the atoms.
The expression~\eqref{eq:FH} can be seen as an augmented version of Dirac's 
delta distribution which is designed such that it can deal
with $r^{-1}$ singularities in the wave function.
We can therefore formally express the Hamiltonian \eqref{eq:Hr} describing 
the relative motion as
\begin{equation}
  \hat{H}_r = \hat{H}_0 + U_0 \ket{O}\bra{O} \label{eq:hr}
\end{equation}
where we define by
\begin{equation}
  \hat{H}_0 = - \frac{\hbar^2}{2 m} \frac{\partial^2}{\partial \vec{r}^2} 
  + \frac{1}{2} m \omega_\perp^2 z^2 \label{eq:H0}
\end{equation}
the noninteracting part of the Hamiltonian and by $\ket{O}\bra{O}$ the 
projector onto the origin in position space, corresponding to the
augmented delta function defined above.
Noting that $U(\sqrt{2}\vec{r}) = U(\vec{r})/\sqrt{8}$ according to 
Eq.~\eqref{eq:FH}, we obtain $U_0 = g_{3D} / \sqrt{8}$.

Owing to the rank--one nature of the perturbation operator in the
expression \eqref{eq:hr} for the Hamiltonian, the Lippmann--Schwinger 
equation describing the scattering process in the relative coordinates 
can be formally solved in terms of the noninteracting retarded Green 
operator $\hat{G}_0(E) = (E - \hat{H}_0 +i0)^{-1}$.
More specifically, we obtain for the full retarded Green operator the
explicit expression
\begin{align}
  \hat{G}(E) & = \left(E - \hat{H}_r +i0\right)^{-1} \nonumber \\
  & = \hat{G}_0(E) + \frac{U_0\hat{G}_0(E)\ket{O}\bra{O}\hat{G}_0(E)}{1 - U_0 \bra{O}\hat{G}_0(E)\ket{O}}
   \,. \label{eq:G}
\end{align}
Its matrix elements in the position representation read
\begin{align}
  \bra{\vec{r}}\hat{G}(E)\ket{\vec{r}'} & = 
  \bra{\vec{r}}\hat{G}_0(E)\ket{\vec{r}'} \label{eq:Gr}  \\
  & \hspace{0.3cm} +\frac{g_{3D} \bra{\vec{r}}\hat{G}_0(E)\ket{\vec{0}} 
    \bra{\vec{0}}\hat{G}_0(E)\ket{\vec{r}'}}{\sqrt{8} - 
    g_{3D} \frac{\partial}{\partial \rho}\left[\rho \bra{\vec{\rho}}
      \hat{G}_0(E)\ket{\vec{0}}\right]_{\vec{\rho}=\vec{0}}} \nonumber 
\end{align}
for $\vec{r}\neq 0$, $\vec{r}'\neq 0$, and $\vec{r}'\neq\vec{r}$, 
where we use the fact that the noninteracting Green function 
$\bra{\vec{r}}\hat{G}_0(E)\ket{\vec{r}'}$ is well--behaved and does not feature 
any singularity for $\vec{r}'\neq\vec{r}$.

The projection of the matrix elements \eqref{eq:Gr} to the 2D plane to which
the atoms are confined gives rise to the equation
\begin{align}
  \bra{\mathbf{r}}\hat{G}(E)\ket{\mathbf{r}'} &  = 
  \bra{\mathbf{r}}\hat{G}_0(E)\ket{\mathbf{r}'} \label{eq:Gr2} \\ 
 & \hspace{0.3cm} + \frac{g_{3D} \bra{\mathbf{r}}\hat{G}_0(E)\ket{\vec{0}} 
    \bra{\vec{0}}\hat{G}_0(E)\ket{\mathbf{r}'}}{\sqrt{8} - 
    g_{3D} \frac{\partial}{\partial \rho}\left[\rho \bra{\vec{\rho}}
      \hat{G}_0(E)\ket{\vec{0}}\right]_{\vec{\rho}=\vec{0}}} \nonumber  \,,
\end{align}
where we define by
$\ket{\mathbf{r}} \equiv \int_{-\infty}^\infty \phi_0(z) \ket{\vec{r}} dz$
the 2D position eigenstate on the plane, which is anchored on the normalized ground
state wavefunction
$\phi_0(z) = (\sqrt{\pi} a_\perp)^{-1/2} \exp[- z^2 / (2 a_\perp^2)]$
of the transverse confinement potential.
Quite straightforwardly, we evaluate
\begin{equation}
  \bra{\mathbf{r}}\hat{G}_0(E)\ket{\mathbf{r}'} =
  \frac{m}{2 i \hbar^2} H_0^{(1)}(k_E |\mathbf{r}-\mathbf{r}'|)
  \label{eq:Hankel2}
\end{equation}
as well as
\begin{equation}
  \bra{\mathbf{r}}\hat{G}_0(E)\ket{\vec{0}} = 
  \bra{\vec{0}}\hat{G}_0(E)\ket{\mathbf{r}}
  = \frac{m}{2 i \hbar^2\sqrt{\sqrt{\pi} a_\perp}} 
  H_0^{(1)}(k_E |\mathbf{r}|) \,, \label{eq:Hankel1}
\end{equation}
with $H_0^{(1)}$ the Hankel function of the first kind of order zero and
\begin{equation}
  k_E = \frac{1}{\hbar}\sqrt{2 m( E - \hbar\omega_\perp/2)}
  \label{eq:kE}
\end{equation}
the in--plane wave number associated with the energy $E$.
The denominator appearing on the right--hand side of Eq.~\eqref{eq:Gr2} was
calculated in Refs.~\cite{PetHolShl00PRL} and \cite{PetShl01PRA},
Assuming that the in--plane kinetic energy of the atoms is much smaller than the
transverse confinement energy $\hbar \omega_\perp$, such that a population of
transversally excited modes is energetically suppressed, this calculation yields
\begin{widetext}
\begin{equation}
  \frac{\partial}{\partial \rho}\left[\rho \bra{\vec{\rho}}
    \hat{G}_0(E)\ket{\vec{0}}\right]_{\vec{\rho}=\vec{0}} =
  - \frac{m}{2\sqrt{\pi}^3 a_\perp \hbar^2} \left[ \ln\left( 
    \frac{2 B}{\pi k_E^2 a_\perp^2} \right) + i\pi \right] \,,
  \label{eq:denom}
\end{equation}
\end{widetext}
with the numerical constant $B \simeq 0.915$ \cite{PetShl01PRA}.
Combining Eqs.~\eqref{eq:g}, \eqref{eq:Gr2}, \eqref{eq:Hankel2}, \eqref{eq:Hankel1},
and \eqref{eq:denom}, we altogether obtain
\begin{align}
  \bra{\mathbf{r}}\hat{G}(E)\ket{\mathbf{r}'} & = 
  \frac{m}{2 i \hbar^2} \Bigg( H_0^{(1)}(k_E |\mathbf{r}-\mathbf{r}'|) \label{eq:rGEr}  \\
  & \hspace{1.6cm}  - \frac{i \pi H_0^{(1)}(k_E |\mathbf{r}|) H_0^{(1)}(k_E |\mathbf{r}'|)}
  {\ln\left(\frac{2 B}{\pi k_E^2 a_\perp^2}\right) + i \pi + \sqrt{2\pi} 
    \frac{a_\perp}{a_s}} \Bigg) \nonumber
\end{align}
for the in--plane position representation of the Green function.
Note that this expression is valid in leading order in $k_E a_\perp$.
Corrections scaling linearly with $k_E a_\perp$ will arise as soon as 
the in--plane kinetic energy of the atoms becomes comparable with the
transverse excitation energy $\hbar\omega_\perp$.


Let us now redo this calculation of the Green function for a
square--lattice discretization of the two--dimensional in--plane space 
with the grid spacing $\delta$.
Using the finite--difference scheme \eqref{eq:fd1} and \eqref{eq:fd2},
we model the projection of the noninteracting part \eqref{eq:H0} of the 
Hamiltonian $\hat{H}_r$ to the transverse ground mode $\phi_0$ as
\begin{align}
  \hat{H}_0^{(\delta)} & = \sum_{l_x,l_y = -\infty}^\infty 
  \left( 2 E_\delta + \frac{\hbar \omega_\perp}{2}\right) 
  \ket{l_x,l_y}\bra{l_x,l_y} \nonumber \\
  & \hspace{0.5cm} - E_\delta \sum_{l_x,l_y = -\infty}^\infty\left( \ket{l_x+1,l_y}\bra{l_x,l_y} \right. \nonumber \\
  & \hspace{3cm}\left. + \ket{l_x,l_y}\bra{l_x+1,l_y} \right) \nonumber \\
  & \hspace{0.5cm} - E_\delta \sum_{l_x,l_y = -\infty}^\infty\left( \ket{l_x,l_y+1}\bra{l_x,l_y} \right. \nonumber \\
  & \hspace{3cm}\left. + \ket{l_x,l_y}\bra{l_x,l_y+1} \right), \label{eq:H0d}
\end{align}
where we define by $\ket{l_x,l_y}$ the localized lattice site orbitals
satisfying $\bracket{l_x,l_y}{l'_x,l'_y} = \delta_{l_xl_x'}\delta_{l_yl_y'}$
and by $E_\delta = \hbar^2/(m\delta^2)$ the characteristic energy scale
of the lattice.
The interaction operator is modeled by a projector onto the origin 
site $\ket{0,0}$ of this lattice according to
\begin{equation}
  \hat{U}_r^{(\delta)} = \frac{U_0^{(\delta)}}{2} \ket{0,0}\bra{0,0} \,.
  \label{eq:Ud}
\end{equation}
$U_{0}^{(\delta)}$ represents the interaction parameter that one would use
in a square--lattice model for the original two--body Hamiltonian \eqref{eq:H}
with the same grid spacing $\delta$, where the interaction operator would 
read 
\begin{equation}
  \hat{U}^{(\delta)} = \sum_{l_x,l_y=-\infty}^{\infty} U_0^{(\delta)} 
  \ket{(l_x,l_y),(l_x,l_y)}\bra{(l_x,l_y),(l_x,l_y)} \,.
\end{equation}
Since this expression is supposed to model the effect of a two--dimensional
delta function and since, as is seen in Eq.~\eqref{eq:Hr}, the prefactor 
$\sqrt{2}$ has to accounted for in the argument of this function when 
transforming from the original particle coordinates to the symmetric 
center--of--mass and relative coordinates \eqref{eq:Rs} and \eqref{eq:rs}, 
we obtain the prefactor $1/2$ in the expression \eqref{eq:Ud} for the 
interaction operator in the relative coordinates. \newline

Since we can write the total lattice Hamiltonian of the relative motion 
projected onto the transverse ground mode as
\begin{equation}
  \hat{H}_r^{(\delta)} = \hat{H}_0^{(\delta)} + \frac{U_0^{(\delta)}}{2} 
  \ket{0,0}\bra{0,0} \,,
\end{equation}
its associated Green operator 
$\hat{G}^{(\delta)}(E) = (E - \hat{H}_r^{(\delta)} + i 0)^{-1}$
is, according to Eq.~\eqref{eq:G}, explicitly expressed as
\begin{equation}
  \hat{G}^{(\delta)}(E) = \hat{G}_0^{(\delta)}(E) + 
  \dfrac{U_0^{(\delta)}\hat{G}_0^{(\delta)}(E)\ket{0,0}\bra{0,0}\hat{G}_0^{(\delta)}(E)}{2 - U_0^{(\delta)} \bra{0,0}\hat{G}_0^{(\delta)}(E)\ket{0,0}}
   \label{eq:GEd}
\end{equation}
in terms of the noninteracting Green operator
$\hat{G}_0^{(\delta)}(E) = (E - \hat{H}_0^{(\delta)} + i 0)^{-1}$.
The latter is calculated via a diagonalization of the noninteracting
lattice Hamiltonian \eqref{eq:H0d} according to
\begin{widetext}
\begin{equation}
  \hat{H}_0^{(\delta)} = \int_{-\pi/\delta}^{\pi/\delta} d k_x 
  \int_{-\pi/\delta}^{\pi/\delta} d k_y \left( \frac{\hbar \omega_\perp}{2} 
  + \left[2 - \cos\left(k_x\delta\right) - \cos(k_y\delta)\right] E_\delta\right)
  \ket{\mathbf{k}}\bra{\mathbf{k}} \,,
\end{equation}
\end{widetext}
with the normalized two--dimensional plane--wave eigenstates 
$\ket{\mathbf{k}} \equiv \ket{k_x,k_y}$ being defined by
\begin{equation}
  \bracket{l_x,l_y}{\mathbf{k}} = \frac{\delta}{2\pi} 
  \exp[i\delta(l_x k_x + l_y k_y)] \,.
\end{equation}
In the continuous limit $\delta\to 0$, which is taken such that
$[(l_x-l'_x)^2 + (l_y-l'_y)^2]^{1/2} \delta$ is kept finite,
we obtain
\begin{equation}
  \bra{l_x,l_y}\hat{G}_0^{(\delta)}(E)\ket{l'_x,l'_y} = 
  \frac{m \delta^2}{2 i \hbar^2} 
  H_0^{(1)}\left(k_E \left|\mathbf{r}_{l_x,l_y}-\mathbf{r}_{l_x',l_y'}\right|\right)
  \label{eq:G0d}
\end{equation}
in perfect analogy with Eq.~\eqref{eq:Hankel2}, where we formally define
$\mathbf{r}_{l_x,l_y} \equiv (l_x\delta,l_y\delta)$.

The diagonal matrix element of the noninteracting Green operator
on the origin site is determined as
\begin{widetext}
\begin{equation}
  \bra{0,0}\hat{G}_0^{(\delta)}(E)\ket{0,0} = 
  \int_{-\pi}^\pi \frac{d\theta}{2\pi} \int_{-\pi}^\pi \frac{d\theta'}{2\pi}
  \frac{1}{E - \hbar \omega_\perp/2 - (2 - \cos \theta - \cos \theta') E_\delta
    + i 0} \,.
\end{equation}
\end{widetext}
Defining
\begin{equation}
  \epsilon = \frac{E - \hbar \omega_\perp/2}{2 E_\delta} = k_E^2 \delta^2 / 4
\end{equation}
with the property $0 < \epsilon \ll 1$ in the continuous limit $\delta\to 0$ \footnote{Strictly speaking, this property is not satisfied in the numerical simulations the results of which are presented in this article, where for the sake of numerical efficiency we chose $k \delta = 1$, which would correspond to $\epsilon = 1$. Numerical convergence checks were performed through comparisons with some test calculations that were carried out for (slightly) lower values of $\delta$.},
where we use the definition \eqref{eq:kE} of the in--plane wave number 
associated with the energy $E$, we obtain through standard residue calculus
\begin{widetext}
\begin{equation}
  \bra{0,0}\hat{G}_0^{(\delta)}(E)\ket{0,0} = 
  \frac{1}{2\pi i E_\delta} \left( \int_0^\epsilon 
  \frac{dx}{\sqrt{x(1-x)(\epsilon-x)(1-\epsilon+x)}}
  - i \int_{\epsilon}^1 \frac{dx}{\sqrt{x(1-x)(x - \epsilon)(1-\epsilon+x)}}
  \right) \,.
\end{equation}
\end{widetext}
Evaluating separately
\begin{eqnarray}
  \int_0^\epsilon \frac{dx}{\sqrt{x(1-x)(\epsilon-x)(1-\epsilon+x)}} & = &
  \pi + \mathcal{O}(\epsilon) \,,\\
  \int_{\epsilon}^1 \frac{dx}{\sqrt{x(1-x)(x - \epsilon)(1-\epsilon+x)}} & = &
  \ln(8/\epsilon) + \mathcal{O}(\epsilon) \hspace{1cm}
\end{eqnarray}
yields the expression
\begin{equation}
  \bra{0,0}\hat{G}_0^{(\delta)}(E)\ket{0,0} = \frac{1}{2\pi E_\delta}
  \left[ \ln(k_E^2 \delta^2/32) - i \pi \right] + \mathcal{O}(k_E^2 \delta^2)
\end{equation}
in the continuous limit.

Inserting this expression and Eq.~\eqref{eq:G0d} into the expression
\eqref{eq:GEd} for the Green operator yields its matrix elements as
\begin{widetext}
\begin{equation}
  \bra{l_x,l_y}\hat{G}^{(\delta)}(E)\ket{l'_x,l'_y} = 
  \frac{m \delta^2}{2 i \hbar^2} 
  \left( H_0^{(1)}(k_E |\mathbf{r}_{l_x,l_y}-\mathbf{r}_{l'_x,l'_y}|) -
  \frac{i \pi H_0^{(1)}(k_E |\mathbf{r}_{l_x,l_y}|) 
    H_0^{(1)}(k_E |\mathbf{r}_{l'_x,l'_y}|)}{\ln
      \left(\frac{32}{k_E^2 \delta^2}\right) + i \pi + 
      \frac{4\pi\hbar^2}{m \delta^2 U_0^{(\delta)}}} \right) \,.
  \label{eq:lGEl}
\end{equation}
\end{widetext}
Comparing this expression with the analogous matrix elements \eqref{eq:rGEr}
of the spatially continuous Green function finally yields the prescription
that we have to choose the square--lattice interaction parameter as
\begin{equation}
  U_0^{(\delta)} = \frac{4\pi \hbar^2 / (m \delta^2)}{\sqrt{2\pi} 
    \frac{a_\perp}{a_s} + \ln\left(\frac{B \delta^2}{16 \pi a_\perp^2}\right)}
  \label{eq:Ud_renorm}
\end{equation}
as a function of the grid spacing $\delta$ in order to obtain a match between
the expressions \eqref{eq:rGEr} and \eqref{eq:lGEl} for the continuous
and discretized Green functions.

In the numerical practice of our calculations, we chose not too fine grids
in order to limit the numerical effort.
Consequently, the logarithmic correction arising in the denominator of
Eq.~\eqref{eq:Ud_renorm} can safely be neglected.
Most specifically, for the choice $\delta = k_E^{-1} = a_\perp$ of the grid spacing
and the inverse wave number, we obtain
$\ln[B \delta^2/(16 \pi a_\perp^2)] \simeq -4$, while we have
$\sqrt{2\pi} a_\perp / a_s \simeq 6 \times 10^2$ for $^{87}$Rb in the presence of the
transverse confinement frequency $\omega_\perp = 2\pi \times 75$~Hz.
This justifies the usage of the ``na\"ive'' expression \eqref{eq:Unaive} for the
interaction parameter in our numerical simulations.

%

\bibliography{bibtex/Julien,bibtex/ComScal,bibtex/GaugeFields,bibtex/AtomLaser,bibtex/ring,bibtex/Peter,bibtex/Peierls,bibtex/CBS,bibtex/AB,bibtex/AndLoc,bibtex/TruncWig,bibtex/Transport,bibtex/graphs,bibtex/NonLinOptics,bibtex/NumTech,bibtex/NonLinOptics,bibtex/NumTech}

\end{document}